\documentclass[preprint,3p,twocolumn]{elsarticle}
\makeatletter
\def\ps@pprintTitle{%
 \let\@oddhead\@empty
 \let\@evenhead\@empty
 \def\@oddfoot{}%
 \let\@evenfoot\@oddfoot}
\makeatother

\usepackage{amssymb}
\usepackage[cmex10]{amsmath}
\usepackage{amsthm}
\usepackage{multicol}
\usepackage{color}
\usepackage{xspace}
\usepackage{adjustbox}
\usepackage{pgf}
\usepackage{csquotes}
\usepackage{etoolbox}
\usepackage[activate={true,nocompatibility},final,tracking=true,kerning=true,spacing=true,stretch=10,shrink=30]{microtype}
\usepackage{url}
\usepackage{amsmath}
\newcommand{\ie}{i.e.}
\newcommand{\etal}{et~al.}
\newcommand{\eg}{e.g.}

\definecolor {infocolor} {rgb} {0.6,0.6,0.6}

\microtypecontext{spacing=nonfrench}

\newcommand{\BigO}{\mathcal{O}}
\newcommand{\whp}{w.h.p.}
\newcommand{\erdos}{Erd{\H o}s-R{\'e}nyi}
\newtheorem{thm}{Theorem}

\newtheorem{lem}[thm]{Lemma}
\newtheorem{cor}[thm]{Corollary}
\newdefinition{rmk}{Remark}
\newproof{pf}{Proof}
\newproof{pot}{Proof of Theorem \ref{thm2}}
\usepackage{amsfonts}

\newcommand{\set}[1]{\left\{ #1\right\}}

\newcommand{\s}{\mathord{-}}

\newcommand{\realrange}[2]{\left[#1, #2\right]}

\newcommand{\unitrange}[2]{\realrange{0}{1}}

\newcommand{\prob}[1]{{\mathbf{P}}\left[#1\right]}

\newcommand{\expect}{{\mathbf{E}}}

\newcommand{\Oh}[1]{\mathcal{O}\!\left( #1\right)}

\newcommand{\Th}[1]{\Theta\!\left( #1\right)}

\newcommand{\Om}[1]{\Omega\!\left( #1\right)}
\newcommand{\Omsmall}[1]{\Omega( #1)}

\newcommand{\llabel}[1]{\label{\labelprefix:#1}}
\newcommand{\labelprefix}{} 

\newcommand{\discussionsize}{\small}

\newcommand{\punkt}{\enspace .}
\newcommand{\komma}{\enspace ,}

\newenvironment{code}{\noindent
\begin{tabbing}%
\hspace{2em}\=\hspace{2em}\=\hspace{2em}\=\hspace{2em}\=\hspace{2em}\=%
\hspace{2em}\=\hspace{2em}\=\hspace{2em}\=\hspace{2em}\=\hspace{2em}\=%
\kill}{\end{tabbing}}

\newcommand{\labelcommand}{}
\newcommand{\captiontext}{}
\newsavebox{\codeparam}
\newcounter{lineNumber}
\newenvironment{disscodepos}[3]{%
\renewcommand{\labelcommand}{#2}%
\renewcommand{\captiontext}{#3}%
\sbox{\codeparam}{\parbox{\textwidth}{#3}}%
\begin{figure}[#1]\begin{center}\begin{code}\setcounter{lineNumber}{1}}{%
\end{code}\end{center}\caption{\llabel{\labelcommand}\captiontext}\end{figure}}

{\end{disscodepos}}

\newdimen\endofsize\endofsize=0.5em
\def\endofbeweis{~\quad\hglue\hsize minus\hsize
                 \hbox{\vrule height \endofsize width
\endofsize}\par}

\newcommand{\CC}{C\raisebox{.08ex}{\hbox{\tt ++}}}

\newcommand{\GG}{g\raisebox{.08ex}{\hbox{\tt ++}}}

\DeclareMathOperator{\acosh}{acosh}

\let\arccos\relax
\DeclareMathOperator{\arccos}{acos}

\newcommand{\defrel}{:=}
\newcommand{\intd}{\text d}
\newcommand{\e}{\ensuremath{e}}
\newcommand{\errorterm}[1]{#1}

\journal{Journal of Parallel and Distributed Computing}

\usepackage{tikz}

\begin{document}

\begin{frontmatter}

\title{Communication-free Massively Distributed\\Graph Generation}

\author[kit]{Daniel Funke} \ead{funke@kit.edu}
\author[kit]{Sebastian Lamm} \ead{lamm@kit.edu}
\author[goethe]{Ulrich Meyer} \ead{umeyer@ae.cs.uni-frankfurt.de}
\author[goethe]{Manuel Penschuck} \ead{mpenschuck@ae.cs.uni-frankfurt.de}
\author[kit]{Peter Sanders} \ead{sanders@kit.edu}
\author[vienna]{Christian Schulz} \ead{christian.schulz@univie.ac.at}
\author[hamilton]{Darren Strash} \ead{dstrash@hamilton.edu}
\author[berlin]{Moritz von Looz} \ead{loozmori@hu-berlin.de}

\address[kit]{Karlsruhe Institute of Technology, Karlsruhe, Germany}
\address[vienna]{University of Vienna, Vienna, Austria}
\address[hamilton]{Hamilton College, Clinton, New York, USA}
\address[goethe]{Goethe University, Frankfurt, Germany}
\address[berlin]{Humboldt University of Berlin, Berlin, Germany}

\begin{abstract}
Analyzing massive complex networks yields promising insights about our everyday lives.
Building scalable algorithms to do so is a challenging task that requires a careful analysis and an extensive evaluation.
However, engineering such algorithms is often hindered by the scarcity of publicly~available~datasets.

Network generators serve as a tool to alleviate this problem by providing synthetic instances with controllable parameters.
However, many network generators fail to provide instances on a massive scale due to their sequential nature or resource constraints.
Additionally, truly scalable network generators are few and often limited in their realism.

In this work, we present novel generators for a variety of network models that are frequently used as benchmarks.
By making use of pseudorandomization and divide-and-conquer schemes, our generators follow a communication-free paradigm.
The resulting generators are thus embarrassingly parallel and have a near optimal scaling behavior.
This allows us to generate instances of up to $2^{43}$ vertices and $2^{47}$ edges in less than 22 minutes on 32\,768 cores.
Therefore, our generators allow new graph families to be used on an unprecedented scale.
\end{abstract}


\end{frontmatter}

\section{Introduction}
Complex networks are prevalent in every aspect of our lives: 
from technological networks to biological systems like the human brain.
These networks are composed of billions of entities that give rise to emerging properties and structures.
Analyzing these structures aids us in gaining new insights about our surroundings.
In order to find these patterns, massive amounts of data have to be acquired and processed.
Designing and evaluating algorithms to handle these datasets is a crucial task on the road to understanding the underlying systems.
However, real-world datasets are often scarce or are too small to reflect future requirement.

Network generators solve this problem to some extent.
They provide synthetic instances based on random network models.
These models are able to accurately describe a wide variety of different real-world scenarios: from ad-hoc wireless networks to protein-protein interactions~\cite{MUTHUKRISHNAN2010686,clauset2009power}.
A substantial amount of work has been contributed to understanding the properties and behavior of these models.
In theory, network generators allow us to build instances of arbitrary size with controllable parameters.
This makes them an indispensable tool for the systematic evaluation of algorithms on a massive scale.
For example, the well known Graph~500 benchmark (\url{graph500.org}), uses the R-MAT graph generator~\cite{chakrabarti2006graph} to build instances of up to $2^{42}$ vertices and $2^{46}$ edges.

Even though generators like R-MAT scale well, the generated instances are limited to a specific family of graphs~\cite{chakrabarti2006graph}.
Many other important network models still fall short when it comes to offering a scalable network generator and in turn to make them a viable replacement for R-MAT.
These shortcomings can often be attributed to the apparently sequential nature of the underlying model or prohibitive hardware requirements.

\subsection*{Our Contribution}
In this work we introduce a set of novel network generators that focus on scalability.
We achieve this by using a communication-free paradigm~\cite{sanders2016scalable}, \ie~our generators require no communication between processing entities (PEs).
An implementation is available as the KaGen library at \url{https://github.com/sebalamm/KaGen}.

Each PE is assigned a disjoint set of local vertices.
It then is responsible for generating all incident edges for this set of vertices.
This is a common setting in distributed computation~\cite{lumsdaine2007challenges}.

The models that we target are the classic \erdos~models $G(n,m)$ and $G(n,p)$~\cite{erdds1959random,gilbert1959random} and different spatial network models including random geometric graphs (RGGs)~\cite{jia2004wireless}, random hyperbolic graphs (RHGs)~\cite{papadopoulos2008greedy} and random Delaunay graphs (RDGs).
The KaGen library also supports the preferential attachment model of Barab{\'a}si and Albert \cite{barabasi1999emergence} using the algorithm from Sanders and Schulz~\cite{sanders2016scalable}.

For each new generator, we provide bounds for their parallel (and sequential) running times.
A key-component of our algorithms is the combination of pseudorandomization and divide-and-conquer strategies.
These components enable us to perform efficient recomputations in a distributed setting without the need for communication. 

To highlight the practical impact of our generators, we also present an extensive experimental evaluation.
First, we show that our generators rival the current state-of-the-art in terms of sequential and/or parallel running time.
Second, we are able to show that our generators have near optimal scaling behavior in terms of weak scaling (and strong scaling).
Finally, our experiments show that we are able to produce instances of up to $2^{43}$ vertices and $2^{47}$ edges in less than 22 minutes.
These instances are in the same order of magnitude as those generated by R-MAT for the Graph~500 benchmark.
Hence, our generators enable the underlying network models to be used in massively distributed settings.

\section{Preliminaries}

We define a \emph{graph} (network) as a pair $G = (V,E)$.
The set $V=\{0,\ldots,n-1\}$ ($|V|=n$) denotes the vertices of $G$.
For a directed graph $E \subseteq V \times V$ ($|E|=m$) is the set of edges consisting of ordered pairs of vertices.
Likewise, in an undirected graph $E$ is a set of unordered pairs of vertices.
Two vertices that are endpoints of an edge $e=\{u,v\}$ are called \emph{adjacent}.
Edges in directed graphs are ordered tuples $e=(u,v)$.
An edge $(u, u) \in E$ is called a \emph{self-loop}.
If not mentioned otherwise, we only consider simple graphs that contain~no~self-loops or parallel edges.

The set of \emph{neighbors} for any vertex $v \in V$ is defined as $N(v) = \{u \in V \mid \{u,v\} \in E\}$. 
For an undirected graph, we define the \emph{degree} of a vertex $v \in V$ as $\text{d}(v) = \Delta(v) = |N(v)|$.
In the directed case, we have to distinguish between the \emph{indegree} and \emph{outdegree} of a vertex.

We denote that a random variable $X$ is distributed according to a probability distribution $\mathcal{P}$ with parameters $p_1,\ldots,p_i$ as $X \sim \mathcal{P}(p_1,\ldots,p_i)$.
The probability mass function of a random variable $X$ is denoted as~$\mu(X)$.
 
\subsection{Network Models}
\subsubsection{\erdos~Graphs}
\label{sec:erdos_related}
The \erdos~(ER) model was the first model for generating random graphs and supports both directed and undirected graphs.
For both cases, we are interested in graph instances without parallel edges. 
We now briefly introduce the two closely related variants of the model.

The first version, proposed by Gilbert~\cite{gilbert1959random}, is denoted as the $G(n,p)$ model.
Here, each of the $n(n-1)/2$ possible edges of an $n$-node graph is independently sampled  with probability $0<p<1$ (Bernoulli sampling of the edges).

The second version, proposed by Erd{\H o}s and R{\'e}nyi~\cite{erdds1959random}, is denoted as the $G(n,m)$ model.
In the $G(n, m)$ model, we chose a graph uniformly at random from the set of all possible graphs which have $n$ vertices and $m$ edges.

For sake of brevity, we only revisit the generation of graphs in the $G(n,m)$ model.
However, all of our algorithms can easily be transferred to the $G(n,p)$ model.

\subsubsection{Random Geometric Graphs}
\label{sec:rgg_model}
Random geometric graphs (RGGs) are undirected spatial networks where we place $n$ vertices uniformly at random in a $d$-dimensional unit cube $[0, 1)^{d}$.
Two vertices $p,q \in V$ are connected by an edge iff their $d$-dimensional Euclidean distance $\text{dist}(p,q) = (\sum_{i=1}^{d}(p_i - q_i)^2)^{1/2}$ is within a given threshold radius $r$.
Thus, the RGG model can be fully described using the two parameters $n$ and $r$.
Note that the expected degree of any vertex that does not lie on the border, \ie~whose neighborhood sphere is completely contained within the unit cube, is $\bar{d}(v) = \pi^{\frac{d}{2}}r^d/\Gamma(\frac{d}{2}+1)$~\cite{penrose2003random}.
In our work we focus on two and three dimensional random geometric graphs, as these are very common in real-world scenarios~\cite{prvzulj2004modeling}.

\subsubsection{Random Hyperbolic Graphs}\label{subsubsec:intro-rhg}
Random hyperbolic graphs (RHGs) are undirected spatial networks generated in the hyperbolic plane with negative curvature.
To generate a RHG, $n$~points are randomly placed on a disk with radius 
\begin{equation}
R = 2 \log n + C,\label{eq:rhg-def-of-R}
\end{equation}
where $C$ controls the average degree $\bar{d}$ with 
\begin{equation}
\bar d = \frac{2}{\pi} [\frac{\alpha}{\alpha - 1/2}]^2 e^{-C/2} (1+o(1))\label{eq:rhg-avg-deg}
\end{equation}
with high probability.~\cite{papadopoulos2008greedy, gugelmann2012random}
Additionally, the model features a dispersion factor $\alpha > 1/2$ affecting concentration of points near the center of the disk.

Each vertex $q$ corresponds to a point with a polar coordinate $\theta_q$ and a radial~coordinate~$r_q$.
Its angle~$\theta_q$ is sampled uniformly at random from the interval $[0, 2\pi)$, while its radius $r_q$ is chosen according to the probability density function
\begin{equation}
    f(r) = \alpha \frac{\sinh(\alpha r)}{\cosh(\alpha R) - 1}. \label{eq:rhg-radial-density}
\end{equation}
Krioukov~\etal~\cite{papadopoulos2008greedy} and Gugelmann~\etal~\cite{gugelmann2012random} show that for $\alpha > 1/2$ the degree distribution in the \emph{threshold model} follows a power-law distribution with exponent $\gamma = 1 + 2\alpha$.
In this RHG variant, two vertices $p,q$ are connected iff their hyperbolic distance 
\begin{multline}\label{eq:hypdist}
  \text{dist}_H(p,q) = \acosh(\cosh r_p \cosh r_q - \\ \sinh r_p \sinh r_q \cos(\theta_p - \theta_q))
\end{multline}
is below the threshold~$R$.
Therefore, the neighborhood of a vertex consists of all the vertices that are within the hyperbolic circle of radius~$R$ around it.

\subsubsection{Random Delaunay Graphs (RDGs)}
A two-dimensional Delaunay graph is a planar graph whose vertices represent points in the plane.
Its edges form a triangulation of this point set, i.e., they partition the
convex hull of the point set into triangles. Furthermore, the
circumcircle of each triangle must not contain other vertices in its
interior. This concept can be generalized for $d$-dimensional
Euclidean space \cite{Edelsbrunner1987}. In particular, for $d=3$ we get a tetrahedralization,
i.e., a decomposition of the space into tetrahedra whose circumsphere
may not contain vertices in their interior.


In this paper, we are concerned with
Delaunay graphs defined by points sampled uniformly at random from the 
$d$-dimensional unit cube $[0,1)^d$ for $d\in\set{2,3}$.
We view this as a good model for
meshes as they are frequently used in scientific computing.
Indeed, these simulations frequently use \emph{periodic boundary conditions},
in order to make small simulations representative for a large simulated system (e.g., \cite{Stukowski2012}).
This can also be viewed as replacing the infinite Euclidean space by a $d$-dimensional torus.
We adopt these periodic boundary conditions, i.e., we implicitly compute the Delaunay-Triangulation of
a point set where for every point $\mathbf{x}$ in the unit cube, also the points
$x+\mathbf{o}$ with $o\in \set{-1,0,1}^d$ are in the point set. Two points in the unit cube
are connected in the output, if any of their copies are connected.
For a scalable distributed graph generator, periodic boundary conditions have the advantage that
we avoid the need to compute some very long edges that appear at the convex hull of random point set.

\subsection{Sampling Algorithms}
\label{subsec:sampling}
Most of our generators require sampling (with/without replacement) of $n$ elements from a (finite) universe $N$.
Sequentially, both of these problems can be solved in expected time $\BigO(n)$~\cite{vitter1987efficient}.
These bounds still hold true, even if a sorted sample has to be generated~\cite{vitter1987efficient,bentley1980generating}.
However, most of these algorithms are hard to apply in a distributed setting since they are inherently sequential.
Recently, Sanders~\etal~\cite{SandersLHSD18} proposed a set of simple divide-and-conquer algorithms that allow sampling $n$ elements on $P$ PEs.
Their algorithms follow the observation that by splitting the current universe into equal sized subsets, the number of samples in each subset follows a hypergeometric distribution.
Based on this observation, they develop a divide-and-conquer algorithm to determine the number of samples for each PE.
In particular, each PE first determines its local interval of the input universe and then recursively generates a set of hypergeometric random variates.
At each level of the recursion, it follows the remaining subset of the universe that contains its local interval.
Hypergeometric random variates are synchronized without the need for communication by making use of pseudorandomization via (high quality) hash functions. 
To be more specific, for each subtree of the recursion, a unique seed value is computed (independent of the rank of the PE).
Afterwards, a hash value for this seed is computed and used to initialize the pseudorandom number generator (PRNG) for the random variates.
Therefore, PEs that follow the same recursion subtrees generate the same random variates, while variates in different subtrees are independent of each other.
Once the remaining subset is smaller than a given threshold, a linear time sequential algorithm~\cite{vitter1987efficient} is used to determine the local samples.
They continue to show that their algorithm runs in time $\BigO(n/P+\log P)$ with high probability\footnote{
\ie~with probability at least $1-P^{-c}$ for any constant $c$
} (\whp)\
if the maximum universe size per PE is $\BigO(N/P)$~\cite{SandersLHSD18}.
Additionally, they demonstrate that their algorithm can be efficiently implemented on Single Instruction Multiple Data (SIMD) architectures,
such as vector units of modern CPUs and general purpose graphic processors (GPGPUs).

We also require the sampling of random numbers that follow a particular probability distribution, \eg~binomial or hypergeometric distributions.
For this purpose, we use the acceptance-rejection method~\cite{robert2013monte, von195113}.
Thus, we are able to generate binomial and hypergeometric random variates in expected constant time $\BigO(1)$~\cite{stadlober1990ratio, stadlober1999patchwork}. 

\subsection{GPGPU Computation Model}
\label{subsec:gpgpu}

The computation and programming model for GPGPUs varies from traditional CPU programming in several aspects.
Computations are organized in blocks of threads.
All threads of a block have access to some common memory block and are able to use synchronization between them.
Blocks, on the other hand, are scheduled independent from each other and have no means of synchronization or communication.
The threads of a block are processed in a SIMD-style manner.
Branches in the code are possible,
however threads of a block taking different branches are no longer processed in parallel but sequentially.

We consider an \emph{accelerator model} where every PE has a GPGPU available to offload computations to but the CPU is considered the main processing  and steering facility.

\section{Related Work}
This paper is the journal version of \cite{FLSSSL18} augmented with more proofs and experiments as well as with material based on the results in \cite{penschuck2017generating}.

We now cover important related work for each of the network models used in our work.
Additionally, we highlight recent advances for other network models that are relevant for the design of our algorithms.

\subsection{ER Model}
Batagelj and Brandes~\cite{batagelj2005efficient} present optimal sequential algorithms for the $G(n, m)$ as well as the $G(n, p)$ model.
Their $G(n,p)$ generator makes use of an adaptation of a linear time sampling algorithm (Algorithm D) by Vitter~\cite{vitter1987efficient}.
In particular, the algorithm samples skip distances between edges of the resulting graph.
Thus, they are able to generate a $G(n,p)$ graph in optimal time $\BigO(n+m)$.
Their $G(n,m)$ generator is based on a virtual Fisher-Yates shuffle~\cite{fisher1938statistical} and also has an optimal running time of $\BigO(n+m)$.

Nobari \etal~\cite{nobari2011fast} proposed a data parallel generator for both the directed and undirected $G(n,p)$ model.
Their generators are designed for graphics processing units (GPUs).
Like the generators of Batagelj and Brandes~\cite{batagelj2005efficient}, their algorithm is based on sampling skip distances but uses pre-computations and prefix sums to adapt it for a data parallel setting.

\subsection{RGG Model}
Generating random geometric graphs with $n$ vertices and radius $r$ can be done na\"ively in time $\Theta(n^2)$.
This bound can be improved if the vertices are known to be generated uniformly at random~\cite{holtgrewe2009scalable}.
To this end, a partitioning of the unit square into squares with side length $r$ is created.
To find the neighbors of each vertex, we consider each cell and its neighbors.
The resulting generator has an expected running time of $\BigO(n + m)$.

Holtgrewe~\etal~\cite{holtgrewe2009scalable, holtgrewe2010engineering} proposed a distributed memory parallelization of this algorithm for the two dimensional case.
Using sorting and vertex exchanges between PEs, they distribute vertices such that edges can be generated locally.
The expected time for the local computation of their generator is $\BigO(n/P \log(n/P))$, due to sorting.
Perhaps more important for large supercomputers is that they need to exchange all vertices resulting in a communication volume of $\Oh{n/P}$ per PE.

We are not aware of efficient distributed implementations of RGG generators for dimensions greater than two. 

\subsection{RHG Model}\label{subsec:related_rhg}
Von Looz \etal~\cite{von2015generating, von2016generating} propose two different algorithms for generating random hyperbolic graphs.
Their first algorithm relates the hyperbolic space to Euclidean geometry using the Poincar\'e disk model to perform neighborhood queries on a polar quadtree.
The resulting generator has a running time of $\BigO((n^{3/2} + m)\log n)$.

In their second approach, von Looz \etal~\cite{von2016generating} propose a generator with an observed running time of $\BigO(n \log n + m)$.
Their algorithm uses a partitioning of the hyperbolic plane into concentric ring-shaped annuli where vertices are stored in sorted order.
Neighborhood queries are computed using angular boundaries for each annulus to bound the number of~vertex~comparisons.

Bringmann \etal~\cite{bringmann2017sampling} introduce a generalization of random hyperbolic graphs called \emph{Geometric Inhomogeneous Random Graphs} (GIRGs).
Their model simplifies theoretical studies of random hyperbolic graphs by ignoring constant factors while maintaining their qualitative behavior.
Additionally, they propose an optimal sampling algorithm for GIRGS with expected linear time.

Finally, independent of this work, Penschuck~\cite{penschuck2017generating} proposed a memory efficient streaming generator that can be adapted to a distributed setting.
Similar to von Looz~\etal~\cite{von2016generating}, they partition the hyperbolic plane into concentric annuli.
They use a sweep-line based algorithm to generate nodes and edges on the fly in a request-centric fashion.
They propose two practical algorithms optimized for either a time complexity of $\BigO(n \log \log n + m)$ or a memory-footprint of $\BigO([n^{1-\alpha}\bar d^\alpha + \log n ]\log n)$ with high probability.
Additionally, they present a shared memory parallelization of their algorithms that can be adapted to a distributed setting with a constant communication overhead.

\subsection{RDG Model}
As the Delaunay triangulation (DT) of a point set is uniquely defined,
generating random Delaunay graphs can be separated into generating a random point set and computing its DT.
A plethora of algorithms for computing the DT of a given point set in two and three dimensions exist.
Funke and Sanders~\cite{Funke2017} review recent work on parallel DT algorithms and 
propose a competitive DT algorithm suitable for large clusters.
The generation of a random point set is identical to the one in the RGG model.

\subsection{Miscellaneous}
\subsubsection{Barabasi and Albert Model}
Batagelj and Brandes~\cite{batagelj2005efficient} give an optimal sequential algorithm
for the preferential attachment model of Barabasi and Albert~\cite{barabasi1999emergence}.
Sanders and Schulz~\cite{sanders2016scalable} parallelize this algorithm that appears to be inherently sequential.
They observe that each edge can be generated independently if the randomness used for generating other edges is reproduced redundantly and consistently using a pseudorandom hash function.
We adapt this technique to other random graph models.



\subsubsection{Recursive Matrix Model}
The recursive matrix model (R-MAT) by Chakrabarti \etal~\cite{chakrabarti2006graph} is a special case of the stochastic Kronecker graph model~\cite{leskovec2005realistic}.
This model is well known for its usage in the popular Graph~500 benchmark.
Generating a graph with $n$ vertices and $m$ edges is done by sampling each of the $m$ edges independently.
For this purpose, the adjacency matrix is recursively subdivided into four partitions.
Each partition is assigned an edge probability $a + b +c + d= 1$.
Recursion continues until a $1 \times 1$ partition is encountered, in which case the corresponding edge is added to the graph.
The time complexity of the R-MAT generator therefore is $\BigO(m \log n)$ since recursion has to be repeated for each edge.

\section{ER Generator}
We now introduce our distributed graph generators, starting with the \erdos~generators for both the directed and undirected case.

\subsection{Directed Graphs}
Generating a (directed) graph in the $G(n,m)$ model is the same as sampling a graph from the set of all possible graphs with $n$ vertices and $m$ edges.
To do so, we can sample $m$ edges uniformly at random from the set of all possible $n (n-1)$ edges.
Since we are not interested in graphs with parallel edges, sampling has to be done \emph{without} replacement.
We do so by using an adaptation of the distributed sampling algorithm by Sanders~\etal~\cite{SandersLHSD18}.

Our generator starts by dividing the set of possible edges into $P$ \emph{chunks}, one for each PE.
Each chunk represents a set of rows of the adjacency matrix of our graph.
We then assign each chunk to its corresponding PE using its id $i$.
Afterwards, PE $i$ is responsible for generating the sample (set of edges) for its chunk.
Note that we can easily adapt this algorithm to an arbitrary number of consecutive chunks per PE.

To compute the correct sample size (number of edges) for each chunk, we use the same divide-and-conquer technique used by the distributed sampling algorithm~\cite{SandersLHSD18} (see Section~\ref{subsec:sampling}).
The resulting samples are then converted to directed edges using simple offset computations.

\begin{thm}
    \label{lem:directedERRunningTime}
    The directed $G(n,m)$ generator runs in time $\BigO((n+m)/{P} + \log P)$ with high probability.
\end{thm}

\begin{pf}
    Our algorithm is an adaptation of the distributed sampling algorithm that evenly divides the set of vertices, and therefore the set of potential edges, between $P$ PEs.
    Thus, the universe per PE has size $\BigO(n(n-1)/P)$ and the running time directly follows from the proof given by Sanders~\etal~\cite{SandersLHSD18}. \qed
\end{pf}

\subsection{Undirected Graphs}
In the undirected case, we have to ensure that an edge $\{i,j\}$ is sampled by \emph{both} PEs, the one that is assigned $i$ and the one that is assigned $j$.
Since each PE is assigned a different chunk, they follow different paths in the recursion tree.
Hence, due to the independence of the random variables generated in each recursion tree, it is highly unlikely that they both sample the edge $\{i,j\}$ independently.
To solve this issue, we introduce a different partitioning of the graphs adjacency matrix~into~chunks.

\begin{figure}[b]
\begin{center}
    \includegraphics[width=0.49\textwidth]{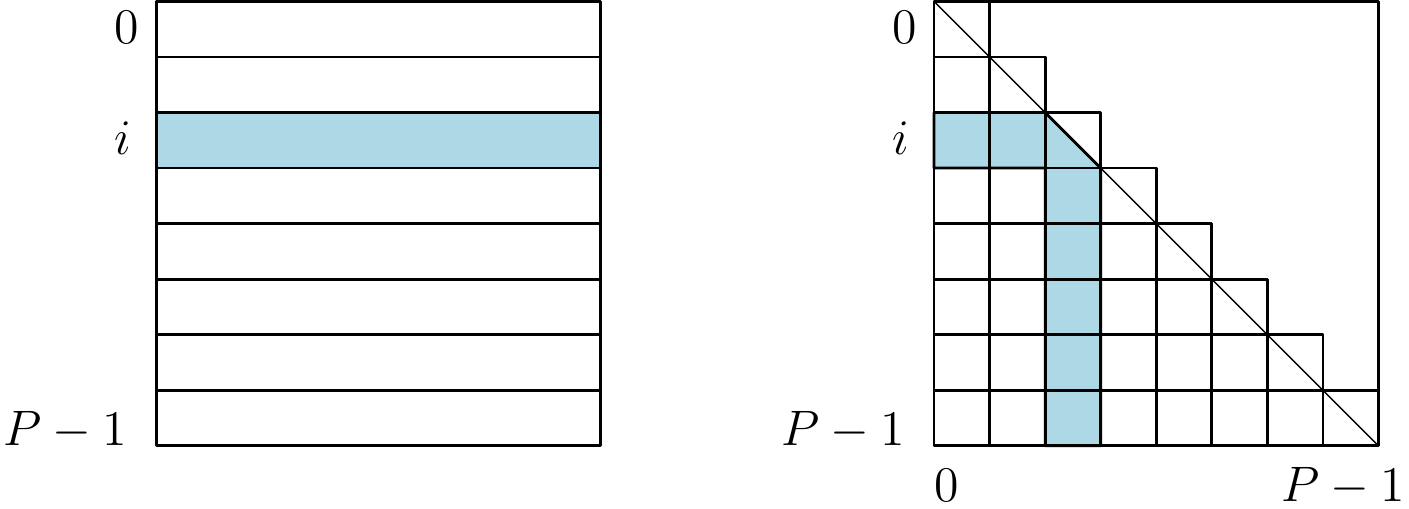}
\end{center}
\caption{\label{fig:chunkMatrix}Examples of an adjacency matrix subdivided into chunks in the directed (left) and undirected (right) case. The chunk(s) for PE $i$ are highlighted in~blue.}
\end{figure}

Our generator begins by dividing each dimension of the adjacency matrix into $P$ sections of size roughly $n/P$.
A chunk is then defined as a set of edges that correspond to a $(n/P) \times (n/P)$ submatrix of the adjacency matrix.
Due to the symmetry of the adjacency matrix, we are able to restrict the sampling to the lower triangular adjacency matrix.
Thus, we have a total of $P(P+1)/2$ chunks that can be arranged into a triangular $P \times P$ chunk matrix.
Afterwards, each PE is assigned a row and column of this matrix based on its id $i$ as seen in Fig.~\ref{fig:chunkMatrix}.
By generating rectangular chunks instead of whole rows or columns, we can make sure that both PE $i$ and PE $j \leq i$ redundantly generate chunk $(i, j)$ using the same set of random values.
In turn, they both sample the same set of edges independently without requiring communication.
Note that our partitioning scheme into chunks results in each chunk being computed twice (once for each associated PE) except for the chunks on the diagonal of our chunk matrix.
Therefore, the recomputation overhead is bounded by $2m$.

We now explain how to adapt the divide-and-conquer algorithm by Sanders~\etal~\cite{SandersLHSD18} for our chunk matrix.
To generate the required partitioning of the adjacency matrix, we start by dividing the $P \times P$ chunk matrix into equal sized quadrants.
This is done by splitting the rows (and columns) into two equal sized sections $\{1,\ldots,l\}$ and $\{l+1,\ldots,P\}$.
We choose $l=\lceil P/2 \rceil$ as our splitting value.


We then compute the number of edges within each of the resulting quadrants.
Since we are only concerned with the lower triangular adjacency matrix, there are two different types of quadrants: triangular and rectangular.
The second and fourth quadrant are triangular matrices with $l$ and $P-l$ rows (and columns) respectively.
We then generate a set of three hypergeometric random variates to determine the number of samples (edges) in each quadrant.
As for the distributed sampling algorithm~\cite{SandersLHSD18}, we make use of pseudorandomization via hash functions that are seeded based on the current recursion subtree to synchronize variates between PEs.

Each PE then uses its id to decide which quadrants to handle recursively.
Note that at each level of the recursion, a PE only has to handle two of the four quadrants.
We use a sequential sampling algorithm once a single chunk remains.
Offset computations are performed to map samples to edges based on the type of the chunk (rectangular or triangular).
The resulting recursion trees has at most $\lceil \log P \rceil$ levels and size $(4P^2 -1)/{3}$.


\begin{thm}
    \label{lem:undirectedERRunningTime}
    The undirected $G(n,m)$ generator runs in $\BigO((n + m)/{P} + P)$ with high probability.
\end{thm}

\begin{pf}
    Each PE $i$ has to generate a total of $P$ chunks consisting of a single triangular submatrix and $P-1$ rectangular submatrices.
    Additionally, each edge $\{i, j\}$ has to be generated twice (except when $P=1$), once by the PE that is assigned vertex $i$, and once by the PE that is assigned vertex $j$.
    Thus, we have to sample a total of $2m$ edges.
    At every level of our recursion, we need to split the quadrants and in turn compute three hypergeometric random variates.
    Therefore, the time spent at every level only takes expected constant time.
    Since there are at most $\lceil \log P \rceil$ levels until each PE reaches its $P$ chunks, the total time spent on recursion is $\sum_{i=0}^{\log P} 2^i = 2(P-1) = \BigO(P)$ with high probability.

    Following the proof by Sanders~\etal~\cite{SandersLHSD18}, we can use Chernoff bounds to show that the total number of samples (edges) that is assigned to any PE will be in $\BigO(m/P)$ with high probability.
    Thus, the undirected $G(n,m)$ generator has a running time of $\BigO((n + m)/{P} + P)$.  \qed
\end{pf}

\subsection{Adaptations for the $G(n,p)$ model}
We now discuss how to adapt our previous generators for the $G(n,p)$ model.
The key observation for the $G(n,p)$ generators is that we do not have to recursively compute hypergeometric random variates in order to derive the correct number of edges for each chunk.
Since the distribution of vertices for each individual chunk is predetermined, we can determine the sample size for each chunk by generating binomial random variates. 
To make sure the sample size for an individual chunk is the same across PEs, the binomial random generator is seeded using a hash value based on the id of the chunk.
Afterwards, we perform the same sampling procedure used to generate edges in the $G(n,m)$ generator.

\subsubsection{Adaption to GPGPUs}
Since the ER generators are a direct application of sampling,
the GPGPU implementation from \cite{SandersLHSD18} can be used to generate graphs on PEs with GPGPUs available.
As before, each PE is assigned a chunk and computes the correct sample size and seeds for the pseudorandom generator on the CPU and then
invokes the GPGPU algorithm to sample the edges of the graph.

\section{RGG Generator}

\label{subsec:rgg_part}
Generating a $d$-dimensional random geometric graph can be done na\"ively in $\Theta(n^2)$ time.
We reduce this bound by introducing a spatial grid data structure similar to the one used by Holtgrewe~\etal~\cite{holtgrewe2009scalable}.
We use a uniform grid of cells with side length $\max(r, n^{-1/d})$.
The vertices of the graph are first placed into the grid cells.
Edges must then run between vertices within one cell or between neighboring cells.
Hence, for a point $A$ assigned to a cell $C$, it suffices to perform distance calculations to the points in cell $C$ and its neighboring cells ($3^d$ cells overall).

%

\begin{thm}
  \label{thm:rggSequential}
  The expected work for generating a random geometric graph with $n$ nodes and $m$ edges is
  $\Oh{n+m}$.
\end{thm}


\begin{pf}
  The work for placing the points is $\Th{n}$.
  The work for initializing the cell array is proportional to
  its size
  $$\left(\frac{1}{\max(r,n^{-1/d})}\right)^d\leq \left(\frac{1}{n^{-1/d}}\right)^d=n\punkt$$
  For estimating the
  remaining work, we estimate the expected number of edges $m$ as well as
  the number of comparisons $Y$ between points.

  Consider the indicator random variable $Z_{ij}$ that is 1 if there
  is an edge between points $i$ and $j$ and 0 otherwise. Then,
  $m=\sum_{i\neq j}Z_{ij}$.  There is an edge between a fixed point
  $i$ and another point $j$ if $j$ is placed in a ball of radius $r$
  (and volume $\Th{r^d}$) around point $i$. Note that at least a
  constant fraction of this ball intersects with the unit cube.  The
  ratio between the volume of this ball and the volume of the unit
  cube is $\Th{r^d}$. Overall, $\prob{Z_{ij}=1}=\Th{r^d}$, and,
  exploiting the linearity of expectation,
  \begin{equation}\label{eq:em}
    \expect m=\sum_{i\neq j}\expect Z_{ij}=\Th{n^2r^d}\punkt
  \end{equation}
  
  Similarly, consider the
  indicator random variable $Y_{ij}$ that is 1 if points $i$ and $j$
  are compared and 0 otherwise. Then, $Y=\sum_{i\neq j}Y_{ij}$.
  Points $i$ and $j$ are compared if they are
  placed into neighboring cells. Recall that each cell has
  $\Th{1}$ neighboring cells (including itself).  
  We now make a case distinction depending on what determines the cell
  size.  If $r\geq n^{-1/d}$ we have $r^{-d}$ cells.  Considering a
  fixed point $i$, a point $j$ is thus placed into one of the $\Th{1}$
  neighboring cells with probability $\Th{r^d}$.  Hence, exploiting the
  linearity of expectation,
  $$\expect Y=\sum_{i\neq j}\expect Y_{ij}=\sum_{i\neq j}\prob{Y_{ij}=1}=\Th{n^2r^d}\punkt$$
  Hence $\expect Y=\Th{\expect m}$.

  When $r<n^{-1/d}$, there are $n$ cells. We can make a similar calculation
  as above, now with $\prob{Y_{ij}}=\Th{1/n}$ which yields $\expect
  Y=n^2/\Th{n}=\Th{n}$.\qed

\end{pf}


\subsection{Parallelization}
We now discuss how to parallelize our approach in a communication-free way.
To the best of our knowledge, the resulting generator is the first efficient distributed implementation of a RGG generator for $d>2$.

Our generator again uses the notion of chunks.
A chunk in the RGG case represents a rectangular section of the unit cube.
We therefore partition the unit cube into $P$ disjoint chunks and assign one of them to each PE.
There is one caveat with this approach, in that the possible values for $P$ are limited to powers of $d$.
To alleviate this issue, we generate more than~$P$ chunks and distribute them evenly between PEs.
To be more specific, we are able to generate $k=2^{db} \geq P$ chunks and then distribute them to the PEs in a locality-aware way by using a Z-order~curve~\cite{morton1966computer}.

Each PE is then responsible for generating the vertices in its chunk(s) as well as all their incident edges.
Again, we use a divide-and-conquer approach similar to the previous~generators.

For this purpose the unit cube is evenly partitioned into $2^d$ equal-sized subcubes.
In turn, the probability for a vertex to be assigned to an individual subcube is the ratio of the area of the subcube to the area of the whole cube.
Thus, we can generate $2^d-1$ binomial random variates to compute the number of vertices within each of the subcubes.
The binomial distribution is parameterized using the number of remaining vertices $n$ and the aforementioned subcube probability $p$.
As for the ER generators, variates are generated by exploiting pseudorandomization via hash functions seeded on the current recursion subtree.
Therefore, we generate the same variates on different PEs that follow the same recursion.
In turn, we require no communication for generating local vertices.
Note that the resulting recursion tree has at most $\lceil \log P \rceil$ levels and size $(2^dP - 1)/{2^d-1}$.
Once a PE is left with a single chunk, we compute additional binomial random variates to get the number of vertices in each cell of side length $c \geq r$.

\begin{figure}[t]
\center
\includegraphics[width=0.32\textwidth]{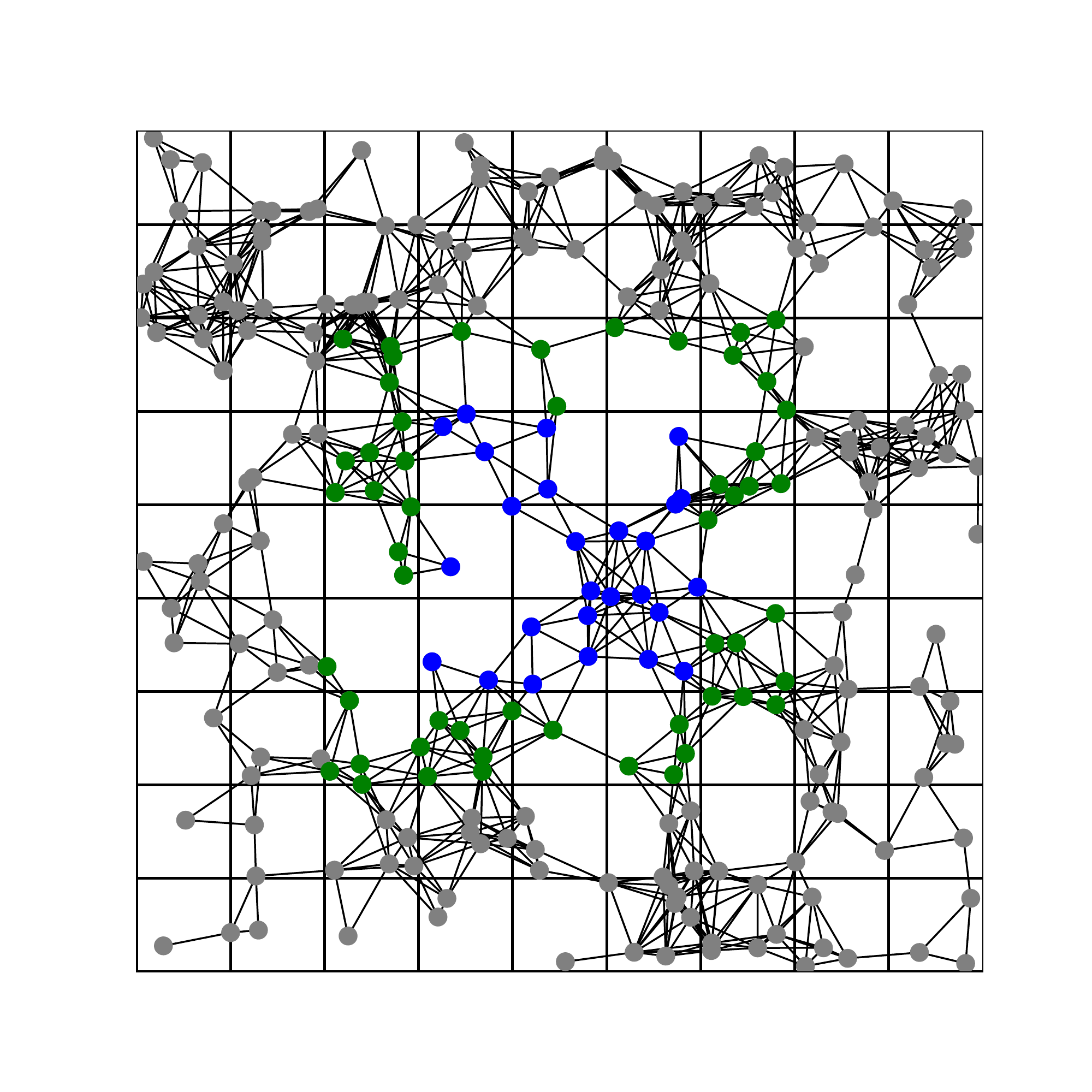}
\caption[Grid data structure]{
    Example of a two dimensional random geometric graph with $256$ vertices and a radius of $0.11$ on nine PEs.
    The local vertices of PE $4$ are highlighted in blue. The non-local vertices computed redundantly by PE 4 are highlighted in green.}
\label{fig:2dRggSingle}
\end{figure}

As we want each PE to generate \emph{all} incident edges for its local vertices, we have to make sure that the cells of neighboring chunks that are within the radius of local vertices are also generated.
Because each cell has a side length $c$ of at least $r$, this means we have to generate all cells directly adjacent to the chunk(s) of a PE.
Due to the communication-free design of our algorithm, the generation of these cells is done through recomputations using the same divide-and-conquer algorithms as for the local cells. 
We therefore repeat the vertex generation process for the neighboring cells.
Note that for sufficiently large graphs, each chunk consists of many cells so that
redundantly generating border layers of cells becomes a negligible overhead.
An example of the subgraph that a single PE generates for the two dimensional case is given in Fig.~\ref{fig:2dRggSingle}.

Afterwards, we can simply iterate over all local cells and generate the corresponding edges by vertex comparisons with all vertices in each neighboring cell.
To avoid duplicate edges, we only perform vertex comparisons for local neighboring cells with a higher id.

\subsection{Analysis of the Parallel Algorithm}
The above communication-free free parallel algorithm emulates a more traditional algorithms that places $n$ points uniformly at random into their cells and performs the necessary distance calculations. Each PE takes time $\Oh{n/P+\log P}$ for generating chunks together with the required parts of the cell array. We do not analyze the number of performed distance calculation directly but indirectly by analyzing the emulated algorithm.
We first note that using standard Chernoff bound arguments, one can prove the following lemmas:
\begin{lem}\label{lem:occupancy}
  If the random variable $\mathrm{Occ}$ denotes the occupancy of a cell then
  $\mathrm{Occ}=\Oh{\expect \mathrm{Occ}+\ln n}$ with probability $1-n^{-c}$ for any constant $c>0$.
\end{lem}
\begin{lem}\label{lem:localNodes}
  If the random variable $\mathrm{W}$ denotes the number of cells
  allocated to one PE then
  $\mathrm{W}=\Oh{n/P+\ln n}$ with probability $1-n^{-c}$ for any constant $c>0$.
\end{lem}

Furthermore, analogous to Theorem~\ref{thm:rggSequential}, one can prove that the
expected work at each PE is $\Oh{(m+n)/P}$ (taking into account that at most a constant fraction of cells has to be generated redundantly).

\begin{lem}\label{lem:exDist}
  The expected number of distance calculations on each PE is $\Oh{(m+n)/P}$.
\end{lem}

However, this does not suffice to bound the parallel execution time since the PE assigned the largest work determines the overall running time.
We conjecture that the amount of work performed on each PE is $\Oh{(m+n)/P}$ \whp for $n=\Om{p\log^2 n}$.
However, we do not know how to prove that formally due to dependencies in the involved random variables (e.g., the variables $Y_{ij}$ and $Z_{ij}$ from the proof of Theorem~\ref{thm:rggSequential}). Instead, we prove the following more loose result.

\begin{thm}
  For any constant $c>0$, there is a constant $a(c)$ such that
  $n\geq a(c)P^2\log^3 P$ implies that
  the amount of work performed by each PE is $\Oh{(m+n)/P}$ with probability at least
  $1-n^{-c}$.
\end{thm}
\begin{pf}
  Let $\mathbf{X}=X_1$,\ldots, $X_n$ denote the vector of positions of the $n$ randomly
  placed points.
  Let
  $Y(\mathbf{X})$ denote the number of distance computations
  performed by a fixed PE.  Since this is a function of $n$ independent
  random variables, we can apply the bounded difference inequality
  \cite{McD89}.
  We have
  \begin{equation}\label{eq:bDiff}
    \prob{Y(\mathbf{X})>\expect Y(\mathbf{X})+\delta}\leq
    \exp\left(-\frac{2\delta^2}{nb^2}\right)\komma
  \end{equation}
  where $b$ is a bound on the maximum change in the value of
  $Y(\mathbf{X})$ when one of the random variables $X_i$ is
  changed. Changing $X_i$ means moving point $i$. This changes the number of distance computation
  by the occupancy of the $\Oh{1}$ cells neighboring the source
  and target cell of the moved point.
  Since the worst case value of the occupancies is very large ($n$), we
  condition on the case that the bound from Lemma~\ref{lem:occupancy} applies. Note
  that the remaining cases are sufficiently unlikely, say have probability $\leq n^{-c}/2$.

  Equation~\ref{eq:bDiff} yields the desired result if
  $\delta$ is large enough such that $\exp(-\frac{2\delta^2}{nb^2})\leq n^{-c}/2P$.
  The factor 2 in the right hand side comes from the fact that we reserve half of the
  allowed failure probability for the above conditioning.
  The factor $P$ comes from the fact that we want to bound the work done on \emph{all} PEs.
  Since we already assume that $n>2P$, we will make the stronger requirement
  $\exp(-\frac{2\delta^2}{nb^2})\leq n^{-(c+1)}$.
  Solving this for $\delta$ yields
  \begin{equation}\label{eq:delta}
    \delta\geq b\sqrt{(c+1)n\ln(n)/2}=\Om{b\sqrt{n\ln n}}\punkt
  \end{equation}

  We now make a case distinction on the ball radius $r$.
  If $r\geq (\ln(n)/n)^{1/d}$, the expected occupancy $\Th{nr^d}$ of a cell is $\Om{\ln n}$ and
  Lemma~\ref{lem:occupancy} yields that $b=\Th{nr^d}$ is also a high probability bound.
  Condition~(\ref{eq:delta})  then becomes
  $\delta=\Omsmall{n^{3/2}r^d\sqrt{\ln(n)}}$.
  At the same time we want $\delta=\Oh{\expect m/P}=\Oh{n^2r^d/P}$; see also Equation~\ref{eq:em}.
  Both conditions can hold if $n^{3/2}r^d\sqrt{\ln(n)}=\Oh{n^2r^d/P}$.
  This is equivalent to $n=\Om{P^2\ln n}$. Since this is only a nontrivial condition when
  $n$ is polynomial in $P$, we get the equivalent condition
  $n=\Om{P^2\ln P}\leq\Om{P^2\ln^3 P}$.

  Similarly, for the case $r< (\ln(n)/n)^{1/d}$, the expected occupancy of a cell
  is $\Oh{\ln n}$ and Lemma~\ref{lem:occupancy}
  yields $b=\Th{\ln n}$. Condition~(\ref{eq:delta}) becomes
  $\delta=\Om{\sqrt{n}\ln^{1.5}n}$.
  We want at the same time that $\delta=\Oh{n/P}$.
  Both conditions hold when $\sqrt{n}\ln^{1.5}n=\Oh{n/P}$,
  or, equivalently, $n=\Om{P^2\ln^3n}$. This
  is equivalent to $n=\Om{P^2\ln^3P}$.\qed
\end{pf}

\subsection{Adaption to GPGPUs}
As before, each PE is responsible for generating the vertices and edges of one chunk.
The algorithm for GPGPUs follows two phases.
In the first phase,
the PE generates the appropriate seeds and vertex numbers for the cells of its chunk and all neighboring cells on the CPU.
Subsequently, the vertices of these cells are sampled on the GPGPU.
Depending on the expected number of vertices per cell,
a cell is either processed by a whole block with several threads or by a single thread,
therefore grouping several cells in one block.
Recall, as the cell side length $c$ is greater than $r$,
only the cells of neighboring chunks immediately adjacent to the PEs chunk need to be generated.

In the second phase,
the actual edges between the vertices are determined,
which requires a three step algorithm.
In the first step,
for each cell and its neighbors, 
the number of edges with length smaller than $r$ are counted on the GPGPU.
Secondly, the prefix sum of these counts provides both the total number of edges generated as well as offsets into the edge array for each block.
The CPU can then allocate memory on the GPGPU for the third stage
and the cells are processed again,
this time actually outputting all edges into the newly allocated array.
The amount of work performed per vertex is the same for all vertices of a cell
 -- as the same number of vertices need to be considered in the neighboring cells --
but can differ between cells.
Therefore, each cell is processed by one block on the GPGPU to avoid any load-balancing issues.

\section{RDG Generator}

The point generation phase for Delaunay graphs follows the same principles as for RGGs,
differing only in the definition of the cell side length~$c$,
which is set to the mean distance of the $(d+1)$th-nearest-neighbor for $n$ vertices in the unit $d$-hypercube,
\mbox{$c\approx( \frac{d+1}{n} )^{1/d}$} \cite{nn}.

To produce the DT of the generated point set, our algorithm proceeds
as follows.  For each assigned chunk, the PE considers the chunk itself plus a \emph{halo}
of neighboring cells.  Initially, the cells directly adjacent to the
chunk are added to the halo.  The PE computes the DT of the chunk plus
halo and checks whether all points of the convex hull are from the
halo and whether each computed simplex $s$ that contains at least one
point from the inside of the assigned chunk has a circumsphere that is completely
contained within the chunk plus halo.  The local computation finishes when
both conditions are fulfilled.  Otherwise, the halo is expanded by one
layer of cells and the DT is updated; see also
\cite{Funke2017,Lo2012}.
As for RGGs, we can ensure that all PEs generate the same vertices for the same cell.

We do not have a complete analysis of the algorithm but note that by
Lemma~\ref{lem:localNodes}, each PE get assigned $\Oh{n/P+\log P}$
nodes from the chunks \whp\ and that the halo contributes only a lower order
term as long as the number of extension steps remains constant. Our
experiments indicate that usually no repetitions at all are needed.
Moreover, a Delaunay triangulation of a random point set can be computed in linear time \cite{Dwyer1991}.
Hence, we might conjecture a running time of $\Oh{n/P+\log P}$ for our algorithm.

\subsection*{Adaptation to GPGPUs}
For the RDG generator, the points can be sampled on the GPGPU according to the algorithm outlined for the RGG generator in the previous section.
Following point generation, the algorithm of Cao et al.~\cite{DelaunayGPU} can be used to compute the DT in two and three dimensions on the GPGPU.
Their algorithm initially produces a near-Delaunay triangulation on the GPGPU and
then fixes potential violations using a star splaying technique on the CPU.
The subsequent steps can be efficiently performed on the GPGPU again:
checking whether the circumhypersphere of all simplices is contained within the halo and,
if not, generating the next layer of halo cells
by first generating the seeds and vertex numbers of those cells on the CPU and then sampling the points on the GPGPU.
Since Cao et al. propose an incremental construction algorithm,
it can be directly applied to insert the newly generated halo points into the DT.

\section{RHG Generators}
We now describe two approaches for generating random hyperbolic graphs.
The first generator (RHG) requires pre-computing local vertices before processing neighborhood queries and allows for an already partitioned output graph.
However, this comes at the cost of load-balancing and memory limitations.
The second generator (sRHG) processes vertices and their incident edges in a streaming fashion.
Although this does not directly give a partitioned output graph, it significantly improves load-balancing and memory requirements.
Additionally, we cover important optimizations that improve the performance of both generators.

\subsection{In-memory Generator}
As for the RGG generator, we can na\"ively create a random hyperbolic graph in $\Theta(n^2)$ by comparing all pairs of vertices.
We first improve this bound by partitioning the hyperbolic plane (cf. \cite{bringmann2015geometric, von2015generating, von2016generating}).
To this end, we split the hyperbolic disk of radius~$R$ into $k := \lfloor {\alpha R}/{\ln(2)} \rfloor$ concentric annuli of constant height, i.e. annulus~$i$ covers the radial interval $[\ell_{i-1}, \ell_i)$ with $\ell_0 = 0$ and $\ell_k = R$.
This results in $k = \Oh{\log n}$ annuli due to Eq.~\ref{eq:rhg-def-of-R}.

Since each PE has to determine the number of vertices in each annulus, we compute a multinomial random variate with $k$ outcomes:
we iteratively compute a set of dependent binomial random variates via pseudorandomization.
The probability that a specific vertex is assigned to annulus~$i$ is given by integration over the radial density function (Eq.~\ref{eq:rhg-radial-density}) between the annulus' limits~$\ell_i$ and~$\ell_{i+1}$:
\begin{equation*}
    p_i = \int_{\ell_i}^{\ell_{i+1}} f(r) dr \stackrel{(Eq.\ref{eq:rhg-radial-density})}{=} \frac{\cosh(\alpha \ell_{i+1}) - \cosh(\alpha \ell_i)}{\cosh(R)-1}.
\end{equation*}

Hence, the expected number $n_i$ of vertices in annulus~$i$ follows $\expect n_i = n \cdot p_i$.

We further partition each annulus in the angular direction into $P$ chunks using a divide-and-conquer approach that uses binomial random variates as for the other generators.
The resulting recursion tree within a single annulus has a height of $\lceil \log P \rceil$.

\begin{figure}[t!]
\begin{center}
    \includegraphics[width=0.4\textwidth]{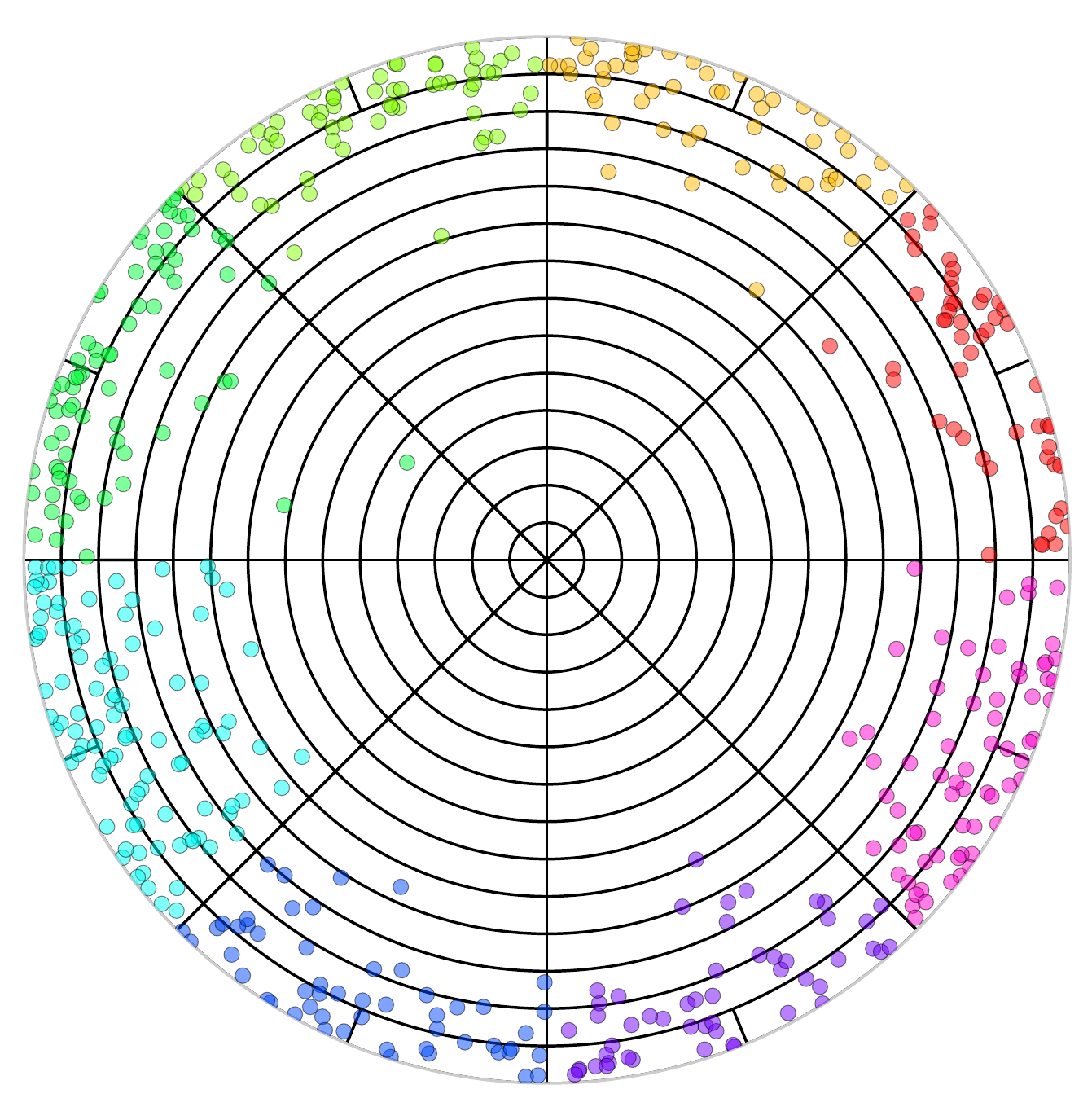}
\end{center}
\vspace*{-8mm}
\caption[Grid data structure]{Partitioning of the hyperbolic plane into set of equidistant annuli and chunks and cells. Each chunk is distributed to one PE and further subdivided into cells. The number of vertices is $n = 512$ with an average degree of $\bar{d} = 4$ and a power-law exponent of $\gamma = 2.6$. The expected number of vertices per cell is set to $2^4$. The local vertices for each PE are highlighted in different colors.}
\label{fig:chunkPartitioning}
\end{figure}

Finally, we perform a third partitioning of chunks into a set of equal-sized cells in the angular direction.
The number of cells per chunk is chosen such that each cell contains an expected constant number of vertices $k$.
Fig.~\ref{fig:chunkPartitioning} shows the resulting partitioning of vertices in the hyperbolic plane into cells and annuli.

\begin{lem}
	\label{lem:rghNodesPerPE}
  Assuming $n = \Omega(P \log P)$, our partitioning algorithm assigns each PE $\Oh{n/P}$ vertices with high probability.
\end{lem}

\begin{pf}
	Chunks are chosen such that they assign each PE $i$ an equally sized angular interval of the hyperbolic plane ${[i \cdot {2\pi}/{P}, (i+1) \cdot {2\pi}/{P})}$.
	The number of vertices per chunk in an annulus is generated through a set of binomial random variates.
	This results in a uniform distribution of the vertices in the interval $[0, 2\pi)$ with respect to their angular coordinate.
	Thus, each PE is assigned $\BigO({n}/{P})$ vertices in expectation.
  Assuming $n = \Omega(P \log P)$ this holds~\whp{} by a standard Chernoff bound.
\end{pf}		

\begin{lem}
    \label{lem:rhgParallelDS}
    Generating the grid data structure for $P$ PEs takes time $\BigO(P \log n + {n}/{P})$ with high probability.
\end{lem}

\begin{pf}
    The time spent during the chunk creation per annulus is $\BigO(P)$~\whp~since the size of the recursion tree is at most $2P-1$ and we only spend expected constant time per level for generating the binomial random variates.
    We have to repeat this recursion for each of the $\BigO(\log n)$ annuli.
    Thus, the total time spent for the recursion over all annuli is $\BigO(P \log n)$.
    The runtime bound then follows by adding the time for vertex creation (Lemma~\ref{lem:rghNodesPerPE}).  \qed
\end{pf}

\subsection*{Neighborhood Queries} 
We now describe how we use our grid data structure to efficiently reduce the number of vertex comparisons.
For this purpose, we begin by iterating over the cells in increasing order from the innermost annulus outwards and perform a neighborhood query for each vertex.

The query begins by determining how far the angular coordinate of a potential neighbor $u = (r_u, \theta_u)$ is allowed to deviate from the angular coordinate of our query vertex $v = (r_v, \theta_v)$.
If we assume that $u$ lies in annulus $i$ with a lower radial boundary $\ell_i$, we can use the distance inequality 
\begin{equation*}
    |\theta_u - \theta_v| \leq \cos^{-1}\Big(\frac{\cosh(r_v)\cosh(\ell_i) - \cosh(R)}{\sinh(r_v)\sinh(\ell_i)}\Big)
\end{equation*}
to determine this deviation.
We then gather the set of cells that lie within the resulting boundary coordinates.
To do so, we start from the cell that intersects the angular coordinate of our query vertex and then continue outward in both angular directions until we encounter a cell that lies outside the boundary.
For each cell that we encounter, we perform distance comparisons to our query vertex and add edges accordingly.
To avoid the costly evaluation of trigonometric functions for each comparison we maintain a set of pre-computed values (see Section~\ref{subsec:opt-trig}).
Note that in order to avoid duplicate edges in the sequential case, we can limit neighborhood queries to annuli that have an equal or larger radial boundary than our starting annulus.

\begin{lem}\label{lem:candidates}
	Consider a query vertex with radius $r$ and an annulus with boundaries $[a, b)$.
	Our candidate selection overestimates the probability mass of the actual query range by a factor of $OE(b{-}a, \alpha)$ where $OE(x,\alpha) \defrel \frac{\alpha - 1/2}{\alpha} \frac{1 - \e^{\alpha x}}{1 - \e^{(\alpha-1/2)x}}$.
\end{lem}

\begin{pf}
	If $r < R{-}b$, the circle around the querying vertex covers the annulus completely.
	Hence, each candidate is a true neighbor and the selection process is optimal.
	
	We now consider $r \ge R{-}a$ and omit the fringe case of $R{-}b < r < R{-}a$ which follows analogously.
	The probability mass $\mu_Q \defrel \mu\left[ B_R(r) \cap (B_b(0) {\setminus} B_a(0)) \right]$ of the intersection of the actual query circle $B_R(r)$ with the annulus $B_b(0) \setminus B_a(0)$ is calculated in Appendix~\ref{eq:mu_actual_query}.
	Our generator overestimates the actual query range at the border and covers the mass $\mu_H\defrel \frac 1 \pi \Delta\theta(r, a) \int_a^b \rho(y) dy$ as detailed in Appendix~\ref{eq:mu_overestimated}.
	The claim follows by the division of both mass functions  ${\mu_H}/{\mu_Q}$.  \qed
\end{pf}

\begin{cor}\label{cor:candidates}
	Given a constant annulus height, i.e. $b{-}a=\BigO(1)$, Lemma~\ref{lem:candidates} implies a constant overestimation for any $\alpha {>} 1/2$.
  In case of $b{-}a=\lfloor {\ln(2)}/{\alpha} \rfloor$, we have $OE(1, \alpha) \le \sqrt{\e} \approx 1.64 \ \ \forall \alpha{>}1/2$.
\end{cor}

\begin{lem}\label{lem:neighbors}
	Let $N_j$ be the number of neighbors the point $p_j {=} (r_j, \theta_j)$ has from below, i.e. neighbors with smaller radius.
	With high probability, \ie~with probability $1-n^{-c}$ for any constant $c>0$, there exist only $\BigO(n / \log^2 n)$ points with $N_j = \BigO(n^{1-\alpha} \bar d ^ \alpha  \log(n))$ while the remainder of points with $r_j > R/2$ has $N_j = \BigO(n^{1 - 2\alpha}\log^{2\alpha}(n) \bar d ^ {2\alpha})$ neighbors.
\end{lem}

\begin{pf}
	Let $X_1, \ldots, X_n$ be indicator variables with $X_i {=} 1$ if $p$ and $p_i$ are adjacent.
	Due to radial symmetry we directly obtain the expectation value of $X_i$ conditioned on the radius $p_i$:
	\begin{align*}
    \expect[{X_i\ |\ r_i{=x}}] & = \prob{X_i {=} 1\ |\ r_i{=x}} \\
                         & = \begin{cases}
		1 & \text{if } x < R - r \\		
		\Delta\theta(x, r) / \pi & \text{otherwise}
		\end{cases} 
	\end{align*}
	
	\noindent We remove the conditional using the Law of Total Expectation and equations (\ref{eq:deltathetaapprox}) and (\ref{eq:cdf}):
	\begin{align*}
    \expect{X_i} &= 
	\int\limits_{0}^{R-r} \rho(x) \intd x 
	\ \ + \ \ \frac 1 \pi \int\limits_{R-r}^r \rho(x) \Delta\theta(x, R) \intd x \\		
  &= \left[\e^{-\alpha r} {-} \e^{-\alpha R}\right](1{+}o(1)) + \frac{1}{\pi}\frac{\alpha }{\alpha {-} \frac 1 2}\e^{-\alpha r} \\ 
  &\phantom{=} 
	\cdot \left[\e^{(\alpha-\frac 1 2)(2r-R)} - 1\right] (1 \pm \BigO(\e^{-r}))\label{eq:cond-exp-val}
	\end{align*}
	
	\noindent
	Fix the radius $r_T  =  R {-} \frac 2 \alpha \log\log n$ with $R/2 < r_T$ (wlog) and consider three cases for $r$:
  First, we ignore all points $r \le R/2$ as they belong to the central clique and are irrelevant here.
  Second, observe that with high probability there exist $\BigO(n / \log^2(n))$ points below $r_T$. 
  Exploiting the monotonicity of Eq.~\ref{eq:cond-exp-val} in $r$, we maximize it by setting $r = R/2$, which cancels out the second term.
  Linearity of the expectation value, substitution of $R = 2\log(n) + C$, and the definition of the expected degree yield
$				\expect{\sum_i X_i} 
				= \BigO\left[n \left({\bar d}/{n} \right)^\alpha \right]  
$.
  Then, Chernoff's bound gives $\sum_i X_i = \BigO(n^{1-\alpha} \bar d ^ \alpha  \log(n))$ with high probability.
	Third, for all points above $r_T$, set $r = r_T$ yielding $\sum_i X_i = \BigO(n^{1 - 2\alpha}\log^{2\alpha}(n) \bar d ^ {2\alpha})$ with high probability analogously.  \qed
\end{pf}

\begin{lem}
    \label{lem:rhgSequential}
    The time complexity of the sequential RHG generator for $n$ vertices with radius $R$, an average degree $\bar d$, and a power-law exponent $\gamma \geq 2$ is $\BigO(m)$ with probability $1-n^{-c}$ for any constant $c>0$.
\end{lem}

\begin{pf}
	\noindent We bound our generators time complexity by considering each component individually:
	\begin{itemize}
		\item The preprocessing requires $\BigO(1)$ time per point making it non-substantial.
		\item Processing the vertices within the cells requires $\BigO(n)$ time in total with high probability.
    \item By applying Lemma~\ref{lem:neighbors} and Cor.~\ref{cor:candidates}, the candidate selection requires $\BigO(n\log \bar d) = \BigO(m)$ time with high probability.
		\item All distance calculations require in total $\BigO(m)$ time since Cor.~\ref{cor:candidates} bounds the fraction of computations that do not yield an edge to $\BigO(1)$.
	\end{itemize}
\end{pf}

To adapt our queries for a distributed setting, we need to recompute all non-local vertices that lie within the hyperbolic circle (of radius $R$) of any of our local vertices.
To find these vertices, we perform an additional inward neighborhood query.

Queries in the distributed setting work similarly to the sequential case, with the addition that any non-local chunks that we encounter during the search are recomputed.
These newly generated vertices are then assigned their respective cells and stored for future searches.

One issue with this approach is that the innermost annulus, which contains only a constant number of vertices (w.h.p.), is divided into $P$ chunks.
However, since all vertices with radius $r \leq R/2$ form a clique and are almost always contained within the search radius for any given vertex, we compute and store these points redundantly in a single chunk on all PEs.
This severely lowers the running time for inward searches, especially for a large number of PEs.

\begin{thm}
   \label{lem:rhgParallel}
   The expected time complexity of the parallel RHG generator for $n$ vertices with radius $R$, average degree $\bar{d}$ and a power-law exponent $\gamma \geq 2$ is ${\BigO(\frac{n + m}{P} + P \log n + n^{1-\alpha} (P\bar{d})^\alpha n^{\frac 1 {2 \alpha}})}$.
\end{thm}

\begin{pf}
	\noindent We bound the time complexity by considering each component individually:
	\begin{itemize}
    \item Building our cell data structures takes time $\BigO(P \log n)$ as shown in Lemma~\ref{lem:rhgParallelDS}.
    \item Sampling local vertices and edges has an expected running time of $\BigO((n+m)/{P})$ as for the sequential approach.
    \item The expected number of vertices recomputed during the inward search can be bounded by $\mu(B_r (0)) = \BigO(n^{1-\alpha} (P\bar{d})^\alpha)$ (see Lemma~\ref{lem:global-vertices}).
    \item The expected number of edges computed for high degree vertices (as well as the number of vertices recomputed during the outward search) can pessimistically be bounded by assuming that vertices in the inner annuli (see Lemma~\ref{lem:global-vertices}) all have maximum degree $\Delta~= ~n^{1 / (2\alpha) + o(1)}$~\cite{gugelmann2012random}. This yields an expected number of $\BigO{(n^{1-\alpha} (P\bar{d})^\alpha n^{1 / (2\alpha)})}$ edges.
\qed
  \end{itemize}
\end{pf}

\subsection{Streaming Generator}
\label{subsec:srhg}
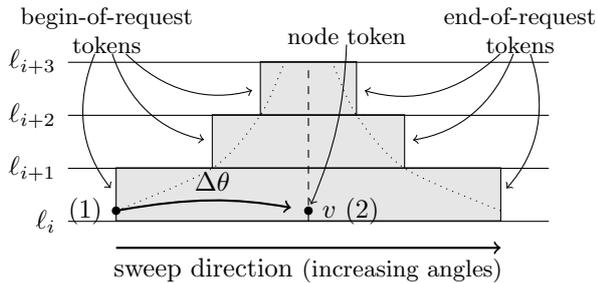
\begin{figure}
    \begin{tikzpicture}[x=1em, y=1em,color=black]
    \def\anheight{2em}
    \def\anwidth{18em}

    \node[anchor=west, inner sep=0, align=center, font=\small] (start-token) at (-0.1*\anwidth, 3.6*\anheight) {begin-of-request\\tokens};
    \node[anchor=east, inner sep=0, align=center, font=\small] (stop-token) at (1.1*\anwidth, 3.6*\anheight) { end-of-request\\tokens};
    \node[inner sep=0, font=\small] (node-token) at (0.58*\anwidth, 3.5*\anheight) {node token};

    \foreach \y/\w in {0/0.8, 1/0.4, 2/0.2} {
        \node[fill=black!10, minimum height=\anheight, minimum width=\w*\anwidth, anchor=south, draw] (region\y) at (\anwidth/2, \y*\anheight) {};
        \path[draw, ->, shorten >=0.3em, bend right] (start-token) to (region\y.west);
        \path[draw, ->, shorten >=0.3em, bend left] (stop-token) to (region\y.east);
    }
    
    \path[draw, dashed] (0.5 * \anwidth, 0) to ++(0, 3*\anheight);

    \foreach \y/\s in {0/, 1/+1, 2/+2, 3/+3} {
        \path[draw, black] (0, \anheight*\y) to (\anwidth, \anheight*\y);
        \node[anchor=east] at (-0.1, \anheight * \y) {$\ell_{i\s}$};
    }
    
    \draw [draw, dotted] plot [smooth] coordinates {
        (0.1 * \anwidth, 0.2 * \anheight)
        (0.3 * \anwidth, 1 * \anheight)
        (0.4 * \anwidth, 2 * \anheight)
        (0.45* \anwidth, 3 * \anheight) };
    
    \draw [draw, dotted] plot [smooth] coordinates {
        (0.9 * \anwidth, 0.2 * \anheight)
        (0.7 * \anwidth, 1 * \anheight)
        (0.6 * \anwidth, 2 * \anheight)
        (0.55* \anwidth, 3 * \anheight) };
    
    \node[draw, circle, inner sep=0.1em, fill, label=left:{(1)}] (pt-sample) at (0.1 * \anwidth, 0.2 * \anheight) {};
    \node[draw, circle, inner sep=0.1em, fill, label=right:{$v$ (2)}] (pt-node) at (0.5 * \anwidth, 0.2 * \anheight) {};
    
    \path[draw, ->, shorten >=0.1em, bend left=10] (node-token) to (pt-node);
    
    \path[->, draw, bend left=10, shorten >=0.5em, thick] (pt-sample) to node[above] {$\Delta\theta$} (pt-node);
    
    \path[->, draw, thick] (0.1 * \anwidth, -0.5 * \anheight) to node[below] {sweep direction {\small (increasing angles)}} (0.9 * \anwidth, -0.5 * \anheight);
    \end{tikzpicture}
    \vspace{-1em}
    \caption{
        The shaded area illustrates the region in which candidates for node~$v$ can be found (an overestimate of the dotted hyperbolic query circle).
        It is encoded with one request per annulus.
        Instead of generating points at random, sRHG draws the \emph{beginning of requests} (1) and then places the points (2) accordingly by increasing their angle by $\Delta\theta$.
        Only when the sweep-line encounters the begin of a request in annulus~$i$, the request is propagated to annulus~$i{+}1$.
    }
    \label{fig:srhg-request}    
\end{figure}

We now present sRHG, a generator that improves the load-balancing and the memory requirements of RHG.
Extending \cite{penschuck2017generating}, its main idea is to invert the neighborhood search:
while RHG selects a node and then directly searches all neighbors by ruling out wrong candidates, sRHG does the opposite and first accumulates all queries a node is candidate in before processing them in a single batch.
This not only reduces unstructured accesses to main memory, but more importantly allows us to narrow the window of space that has to be kept in memory.
 
As for RHG, we decompose the hyperbolic plane into a set of concentric annuli and draw the number of points in each annuli uniformly at random.
This step is performed by all PEs independently and we use pseudorandomization to ensure that each PE draws the same numbers without communication.

Next we perform the second decomposition by splitting each annulus in the angular direction into a set of $P$ chunks of equal size.

Conceptually, sRHG then executes a sweep-line algorithm in angular direction starting with the innermost annulus.
To this end, the PE maintains a sweep-line state per annulus storing the currently active requests and pending events:
as illustrated in Fig.~\ref{fig:srhg-request}, we use tokens for each node~$v$ to mark the \emph{position} of~$v$, as well as the \emph{begin} and \emph{end} of the region in which neighbor candidates of~$v$ can be found.

sRHG executes the next pending event (i.e. the unprocessed token with smallest polar coordinate) of the current annulus:
\begin{itemize}
    \item A \emph{begin-of-request} token adds the request to the current sweep state, and creates a \emph{begin-of-request} token for the next higher annulus (if existing). It also generates an appropriate \emph{end-of-request} token for the current annulus.
    \item An \emph{end-of-request} removes the request from the current sweep state.
    \item If a \emph{node token} is found, the distance to each node with a request in the current sweep state is computed and an is edge emitted if it is below the threshold~$R$.
\end{itemize}

Observe this design causes sRHG to process the begin-of-request token of a node~$v$ \emph{before} its node token becomes active.
We hence invert the causality relation, and draw begin-of-request tokens from a monotonic sequence of uniform variates and position a matching point token accordingly.

To further reduce the memory footprint, we do not complete an annulus before starting the next higher one.
Instead, each PE interleaves the processing of its local annuli with the only constraint that no sweep-line may overtake the sweep-line below it.
This is legal, since the only information flow is from lower to higher annuli: it is triggered by begin-of-request tokens which never move towards smaller angles during this process.

\begin{figure}[t]
  \centering
	{\adjustbox{trim={.24\width} {.4\height} {0.0\width} {.0\height},clip}{
		\scalebox{0.8}{\input{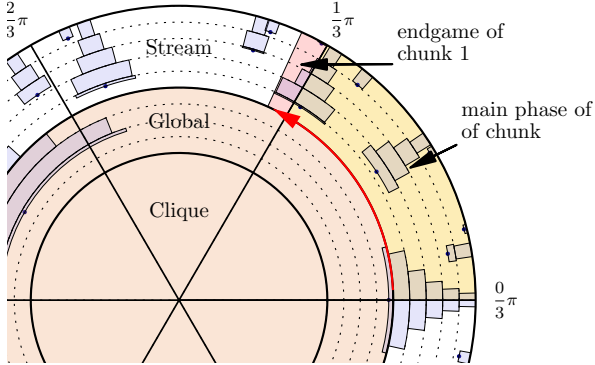}}
	}}
	\caption{\label{fig:hypgen}
		The hyperbolic plane is partitioned along the polar axis into $P$ chunks of equal size.
		Radially, there are two groups: the \emph{lower global annuli} which are preprocessed and kept in memory, and the \emph{upper streaming annuli}.
		In the main phase, each PE streams through its chunks towards increasing polar angles (red arrow).
    Requests overlapping into the next chunk are then completed in the \emph{final phase}.
  }
	
\end{figure}

After the decomposition, sRHG partitions the annuli into two groups, \emph{lower global annuli} and \emph{upper streaming annuli}, and starts by processing the global annuli first (see Fig.~\ref{fig:hypgen}):
\begin{itemize}
    \item \emph{Global annuli} are those where the maximum potential request width of a point within that annulus is larger than $2\pi/P$ (the width of a single chunk). 
    Similarly to RHG, we merge the innermost annuli with a radial boundary below $R/2$ into a special clique annulus.
    \item \emph{Streaming annuli} are those where the maximum potential request width of a point within that annulus is at most $2\pi/P$.
\end{itemize}

Let $r_G$ be the smallest radial boundary of a streaming annulus (i.e. every streaming annulus~$i$ has a lower boundary of $\ell_{i-1} \ge r_G$).
We obtain $r_G \lesssim R/2 + \log (2P / \pi)$ for $P \ge 2$ by solving $2\Delta\theta(r_G, r_G) = {2\pi}/{P}$ for $r_G$ and applying Eq.~\ref{eq:deltathetaapprox}.

\begin{lem}\label{lem:global-vertices}
  The expected number of vertices generated in the global annuli is $\BigO(n^{1-\alpha} (P \bar d)^{\alpha})$.
\end{lem}

\begin{pf}
    Consider a point $(r_G, \theta)$ with a request width of at most $2\Delta\theta(r_G, r_G)$.
    The number of vertices $n_G(P)$ that each PE has to generate during the global phase is thus binomially distributed around the mean of
    \begin{align*}
    \expect{n_G(P)} 
    & = n\mu(B_{r_G}(0)) 
    = n \left(\frac{\bar d P}{2n}\right)^\alpha \left(\frac{\alpha - \frac 1 2}{\alpha}\right)^{2\alpha} \\
    & = \BigO\left(n^{1-\alpha} (P \bar d)^\alpha \right).
    \end{align*}
    \qed
\end{pf}

\begin{lem}\label{lem:stream-vertices}
  Assuming $n = \Omega(P \log P)$, the number of vertices generated in the streaming annuli of any PE is $\BigO(n / P)$ with high probability.
\end{lem}

\begin{pf}
    Each PE is assigned a polar interval of $2\pi / P$ width and generates all streaming points whose request begins there.
    As the angle at which a request starts is drawn uniformly at random, the numbers $(n_1, \ldots, n_P)$ of vertices generated by each PE are distributed multinomially with an equal bucket mass of $p = 1/P$.
    We pessimistically place all point with in the streaming annuli (cf. Lemma~\ref{lem:global-vertices}), and hence have $\sum_i n_i = n$ and $\expect{n_i} = \BigO(n / P)$ for all $i$.
    Concentration follows directly from Chernoff bounds. \qed
\end{pf}

\subsection*{Global annuli} 
By construction, points in the global annuli have long requests, potentially covering the whole hyperbolic space.
In order to guarantee that no PE has to generate all vertices, requests within the lower global annuli are computed redundantly on all PEs.
Again, consistency across PEs is achieved using pseudorandomness.
Each PE then restricts the requests to its own streaming chunk and propagates applicable ones to a designated insertion buffer in the first upper streaming annulus.

During the creation of a request, it might happen that we encounter an angular deviation of $[a,b]$ where either $a < 0$ or $b > 2\pi$.
Taking the angular $2\pi$-period of the hyperbolic plane into account, these requests are separated into two ranges.
To be more specific, we separate the angular deviation $[a,b]$ with $a<0$ into the two ranges $[a + 2\pi, 2\pi]$ and $[0,b]$.
The case $b>2\pi$ is treated analogously.

Due to the nature of hyperbolic space, vertices in the global annuli are likely to have a very high degree as they have a relatively small hyperbolic distance to any other point.
However, due to our request-centric approach the computation of their neighbors in the upper streaming annuli is fully distributed.

To achieve a good scaling, we distribute the execution of requests for the inner annuli evenly across all PEs.
This results in a much more even distribution of work compared to the query-centric approach.
However, it does not directly produce a partitioned output graph.

\subsection*{Streaming annuli} 
After this so-called global phase, we continue with the upper streaming annuli.
By construction, each PE is responsible for generating and processing all requests within its local chunks (i.e. one per annulus).
It maintains a context for each of its streaming annuli consisting of:
\begin{itemize}
    \item The sweep state containing all active requests.
    \item An PRNG emitting monotonically increasing variates distributed uniformly over the chunk's polar interval.
    \item A priority queue to receive \emph{begin-of-request} tokens from the annuli below (referred to as insertion buffer).
    \item A priority queue storing points generated after drawing their \emph{begin-of-request}.
    \item A priority queue storing \emph{end-of-request} tokens.
\end{itemize}

As aforementioned, we execute a sweep-line algorithm and always process (and remove) the token with smallest angle from one of the four sources.
Note that by definition of the angular width of each request, the candidate selection for each request gives the same overestimation as our previous generator (see Lemma~\ref{lem:rhgSequential}).
The in-order generation of requests and edges however significantly decreases unstructured memory accesses compared to RHG.

While sRHG needs to process annuli in an interleaved fashion to bound the insertion buffers' sizes, it tries to infrequently switch between annuli to improve data locality.
We implemented this by splitting each chunk into cells; the number of cells per annulus is chosen such that the expected number of points in them is constant.
The algorithm then switches between annuli only after processing complete cells.

We conclude the main generation if all sweep-lines reached the upper bound of the PE's polar interval.
Observe that at this point, unprocessed \emph{node} or \emph{end-of-request} token can remain which are dealt with during final phase.

For the final phase, each PE is responsible for generating edges from pairs where either the request or vertex stem from the main phase. 
To do so, we repeat the same process as in the local phase, but replicate the configuration of the PE responsible for the adjacent chunks.
We also annotate vertices and requests from these foreign chunks in order to not generate duplicate edges.

The final phase involves at most one additional chunk as all points and requests with large polar shift were processed during the global phase.
In practice, it can often be stopped earlier once all old vertices and requests are processed, we count their number.

\begin{lem}\label{lem:stream-edges}
  If we limit the final phase to the size of a chunk, the expected number of edges generated for streaming annuli is $\BigO((\frac n P)^{2-\alpha} \bar d^\alpha)$.
\end{lem}

\begin{pf}
    Following the proof of Lemma~\ref{lem:neighbors}, we introduce indicator variables $X_1, \ldots, X_n$ with $X_i = 1$ if two points $p$ and $p_i$ are adjacent. 
    This yields the expected value
    \begin{align*}
    \expect{X_i} &= 
    \int\limits_{0}^{R-r} \rho(x) \intd x 
    \ \ + \ \ \frac 1 \pi \int\limits_{R-r}^r \rho(x) \Delta\theta(x, R) \intd x \\		
    &= \left[\e^{-\alpha r} {-} \e^{-\alpha R}\right](1{+}o(1))
    + \frac{1}{\pi}\frac{\alpha }{\alpha {-} \frac 1 2}\e^{-\alpha r} \\
    &\phantom{=} 
    \cdot \left[\e^{(\alpha-\frac 1 2)(2r-R)} - 1\right] (1 \pm \BigO(\e^{-r}))
    \end{align*}
    We now (pessimistically) assume $r = r_G$ (Lemma~\ref{lem:global-vertices}).
    The resulting expected number of neighbors is given by $\expect{X_i} = \BigO(P^{\alpha - 1} (\bar{d} / n)^\alpha - (\bar{d} / n)^{2 \alpha})$.
    In turn, we can use Lemma~\ref{lem:stream-vertices} to bound the expected total number of request and vertex pairs for vertices in the streaming annuli by 
    $
    \BigO((n / P)^{2-\alpha} \bar d^\alpha).
    $ \qed
    
\end{pf}

\begin{lem}\label{lem:sequential-rt}
  The time complexity of the sequential sRHG generator for $n$ vertices with radius $R$, an average degree $\bar d$, and a power-law exponent $\gamma \geq 2$ is $\BigO(m)$ with probability $1-n^{-c}$ for any constant $c>0$.
\end{lem}

\begin{pf}
    \noindent We bound the time complexity by considering each component individually:
    \begin{itemize}
        \item The preprocessing requires $\BigO(1)$ time per point making it non-substantial.
        \item Handling of cliques is trivially bounded by $\BigO(m)$ since every iteration emits an edge.
        \item By applying Lemma~\ref{lem:neighbors} and Corollary~\ref{cor:candidates}, the candidate selection requires $\BigO(n\log \bar d) = \BigO(m)$ time with high probability.
        Here we exploit that request tokens can be addressed to discrete cells allowing for linear time integer sorting.
        \item All distance calculations require in total $\BigO(m)$ time since Cor.~\ref{cor:candidates} bounds the fraction of computations that do not yield an edge to $\BigO(1)$. \qed
    \end{itemize}
\end{pf}

\begin{thm}
   \label{lem:parallel-rt}
   The expected time complexity of the parallel sRHG generator for $n$ vertices with radius $R$, average degree $\bar{d}$ and a power-law exponent $\gamma \geq 2$ is ${\BigO(\frac{n+m}{P} + P \log n + n^{1-\alpha} (P \bar{d})^\alpha + (\frac n P)^{2-\alpha} \bar d^\alpha)}$.
\end{thm}

\begin{pf}
    \noindent We bound the time complexity by considering each component individually:
    \begin{itemize}
        \item Building our cell data structures takes time $\BigO(P \log n)$ as shown in Lemma~\ref{lem:rhgParallelDS}.
        \item The expected number of vertices that have to be recomputed for the global annuli $\BigO(n^{1-\alpha} (P\bar{d})^\alpha)$ due to Lemma~\ref{lem:global-vertices}.
        \item Lemma~\ref{lem:stream-edges} bounds the expected number of distance comparisons for the outer annuli to $\BigO \left( (n/P)^{2-\alpha} d^\alpha \right)$. \qed
    \end{itemize}
\end{pf}

Given the running time of our parallel algorithm, we can now assume $n/P = k = \Omega(\log n)$ vertices per PE and set $k$ to $n^{2/3}$.
For values of $\gamma \geq 3$ this results in a running time linear in the number of edges per PE $\BigO(m/P)$.
However, for very small values of $\gamma$ close to $2$, the running time is dominated by the global phase an becomes nearly linear.

\subsubsection{Further Optimizations} 
\paragraph{Adjacency tests without trigonometric functions}\label{subsec:opt-trig}
The runtime of preliminary versions of our generators was dominated by repeated evaluations of trigonometric functions.
We address this issue with a pre-computing scheme.
Let $p = (r_p, \theta_p)$ and $q = (r_q, \theta_q)$ be two arbitrary vertices in $\mathcal{D}_R$ and let $\ell_i$ be the lower radial boundary of annulus~$i$.

The scheme accelerates the two most frequent query types, namely the computation of request widths of the form $\Delta\theta(r_p, \ell_i)$ and tests whether the hyperbolic distance $d(p, q) < R$ of two vertices falls below the threshold $R$.
The former computation type occurs $\Omega(n)$ times, while the latter is required in each of the $\Omega(m)$ candidate checks.
Based on Equation~\ref{eq:deltatheta} we obtain
\begin{multline}
	\Delta\theta(r_p, \ell_i) = \cos^{-1}\Big(\coth(r_p) \cdot \coth(\ell_i)\\
		- \frac{\cosh(R)}{\sinh(\ell_i)} \cdot \frac{1}{\sinh(r_p)} \Big),
\end{multline}
where $\coth(x) := \cosh(x) / \sinh(x)$.
By pre-computing $\coth(r_p)$ and $1 / \sinh(r_p)$ for each vertex, as well as $\coth(\ell_i)$ and $\cosh(R) / \sinh(\ell_i)$ for each annulus, the argument of $\cos^{-1}$ follows from two multiplications and a subtraction.

Similarly, we test $d(p, q) < R$ by rewriting Equation~\ref{eq:hypdist}:
\begin{multline}
	\label{eq:effdistcomp}
	\cos(\theta_p)\cos(\theta_q) + \sin(\theta_p)\sin(\theta_q) > \\ 
	\coth(r_p) \coth(r_q) - \cosh(R) \frac{1}{\sinh(r_p)} \frac{1}{\sinh(r_q)},
\end{multline}
The left-hand-side is an expansion of $\cos(\theta_p - \theta_q)$.
Hence, the additional pre-computation of $\cos(\theta_p)$ and $\sin(\theta_p)$ for each vertex gives distance checks in at most five multiplications and two additions, which can be further improved by re-using parts of earlier computations.

After pre-computation, Expression~\ref{eq:effdistcomp} can be vectorized efficiently to compute the distance between a node and multiple requests in a data-parallel fashion.
To support vectorized computations, we also use a structure-of-arrays memory layout to store active candidates.
We employ the Vc library \cite{DBLP:journals/spe/KretzL12} for explicit vectorization.

\paragraph{Batch-processing of requests}
Our streaming generator effectively sweeps all annuli in an interleaved fashion and maintains for each annulus a separate state containing the active candidates.
During this sweep, it encounters three event types, namely the occurrence of a vertex, the beginning of a request, and eventually its end.

Our implementation splits each annulus into cells of equal width and then processes these events batch-wise.
Given the number $n_j$ of vertices in annulus~$j$, we select the number $c_j$ of cells in annulus~$j$ such that $c \le n_i / c_i < 2c$ where $c$ is a small tuning parameter (typically $8$).
More precisely, we choose $c_j$ as a power-of-two which by construction aligns cell boundaries between annuli and avoids corner cases when traversing the geometry.

When entering a cell, we move all requests contained into the active state by first overwriting obsolete requests that went out-of-scope in the last cell; we thereby avoid redundant operations otherwise caused by separated deletions and insertions.
Subsequently, all vertices contained are matched against the active candidates again increasing data locality and exploiting minor synergies.

The usage of cells also allows us to relax the sorting of requests received from below since we only need to distribute start and end-points of requests to the appropriate cells.
Our implementations hence only stores for each request the indices of cells in which it starts and ends, and orders the items in a radix heap.

\section{Experimental Evaluation}
We now present the experimental evaluation of our graph generators.
For each algorithm, we perform a running time comparison and analyze its scaling behavior.

\subsection{Implementation}
An implementation of our graph generators (KaGen) in \CC~is available at \url{https://github.com/sebalamm/KaGen}.
We use Spooky Hash\footnote{\url{http://www.burtleburtle.net/bob/hash/spooky.html}} as a hash function for pseudorandomization.
Hash values are used to initialize a Mersenne Twister~\cite{matsumoto1998mersenne} for generating uniform random variates.
Non-uniform random variates are generated using the stocc library\footnote{\url{http://www.agner.org/random/}}.
If the size of our inputs (\eg~the adjacency matrix size) exceeds $64$ bit, we use the multiple-precision floating points library GMP\footnote{\url{http://www.mpfr.org}} and a reimplementation of the stocc library.
Profiling indicates that most generators spend only a negligible fraction of their time in random number generation ($\leq 1$~\%). For the ER generator this figure is about 20~\%. Hence, we did not experiment with alternative implementations.
All algorithms and libraries are compiled using \GG~version 5.4.1 using optimization level \texttt{fast} and \texttt{-march=native}.
In the distributed setting, we use Intel MPI~version 1.4 compiled with \GG~version 4.9.3.

\subsection{Experimental Setup}
\label{sec:exp_setup}
We use two different machines to conduct our experiments.
Sequential comparisons are performed on a single core of a dual-socket Intel Xeon E4-2670 v3 system with 128 GB of DDR4-2133 memory, running Ubuntu 14.04 with kernel version 3.13.0-91-generic.
If not mentioned otherwise, all results are averages of ten iterations with different seeds.

For scaling experiments and parallel comparisons we use the Phase 1 thin nodes of the SuperMUC supercomputer.
The SuperMUC thin nodes consist of 18 islands and a total of 9216 nodes.
Each compute node has two Sandy Bride-EP Xeon E5-2680 8-core processors, as well as 32 GB of main memory.
Each node runs the SUSE Linux Enterprise Server (SLES) operating system.
We use the maximum number of 16 cores per node for our scaling experiments.
The maximum size of our generated instances is limited by the memory per core (2 GB).
To generate even larger instances, one could use a full streaming approach which will be discussed in Section~\ref{sec:future}.

We analyze the scaling behavior of our algorithms in terms of weak and strong scaling.
Weak scaling measures how the running time varies with the number of PEs for a fixed problem size \emph{per PE}.
Analogously, strong scaling measures the running time for a fixed problem size over \emph{all PEs}.
Due to memory limitations of the SuperMUC, strong scaling experiments are performed with a minimum of 1024 PEs.
Again, results are averaged over ten iterations with different seeds.

\subsection{\erdos~Generator}
There are various implementations of efficient sequential \erdos~generators available (\eg~\cite{batagelj2005efficient}).
However, there is little to no work on distributed memory generators.
Thus, we perform a sequential comparison of our generator to the implementation found in the Boost\footnote{\url{http://www.boost.org/doc/libs/1_62_0/libs/graph/doc/erdos_renyi_generator.html}} library.
Their generator uses a sampling procedure similar to Algorithm~D~\cite{vitter1987efficient} and serves as an example for an efficient linear time generator.

\begin{figure}[t]
  \centering
  \includegraphics[width=0.49\textwidth]{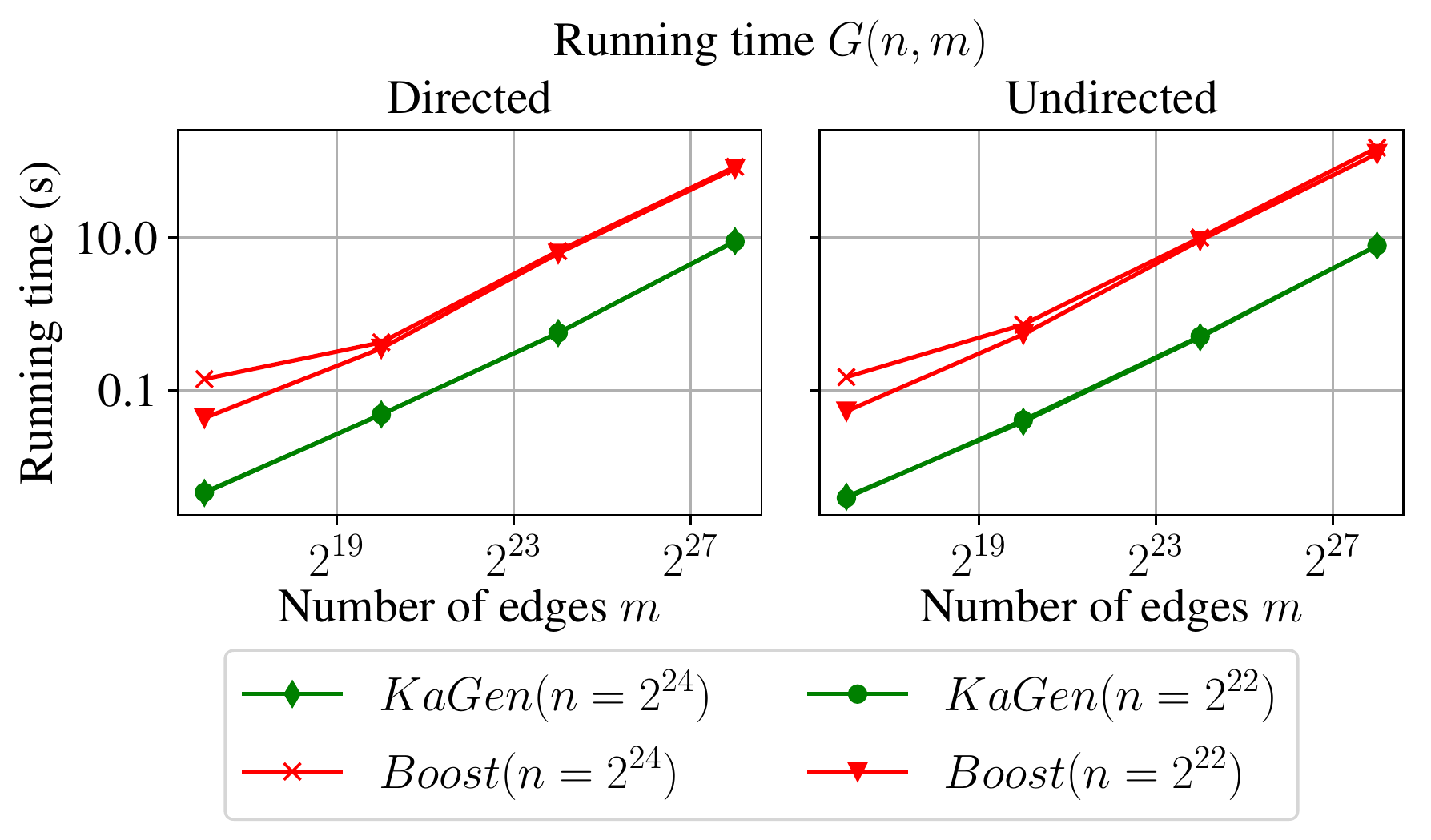}
  \vspace*{-.7cm}
  \caption{\label{fig:er_seq}Running time for the sequential directed (left) and undirected (right) \erdos~generators for $2^{22}$ and $2^{24}$ vertices and $2^{16}$ to $2^{28}$ edges.}
\end{figure}

For our comparison, we vary the number of vertices from $2^{18}$ to $2^{24}$ and the number of edges from $2^{16}$ to $2^{28}$.
Fig.~\ref{fig:er_seq} shows the running time for both generators for the two largest sets of vertices.

First, we note that both implementations have a constant time per edge for large $m$.
However, the Boost implementation also has an increasing time per edge for growing numbers of vertices $n$.
In contrast, the running time of our generator is independent of $n$.
This is no surprise, since our generator uses a simple edge list and does not maintain a full graph~data~structure.

For the directed $G(n,m)$ model, our generator is roughly $10$ times faster than Boost for the largest value of $m=2^{28}$.
In the undirected case, our $G(n,m)$ generator is roughly $21$ times faster and has an equally lower running time
All in all, the results are consistent with the optimal theoretical running times of $\BigO(n+m)$ for both algorithms.

\begin{figure}[t]
  \centering
  \includegraphics[width=0.49\textwidth]{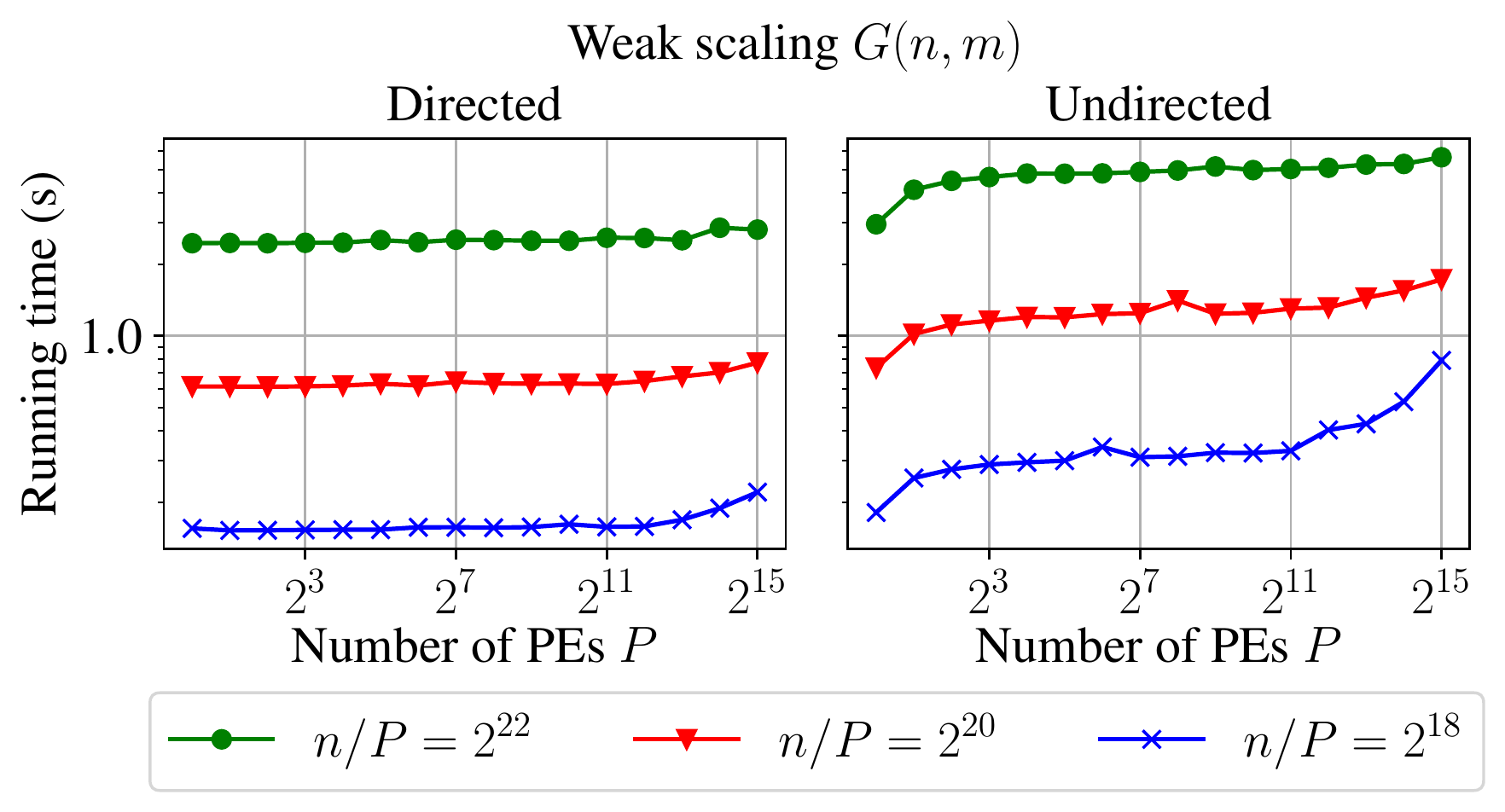}
  \vspace*{-.7cm}
  \caption{\label{fig:gnmPar}Running time for generating $m$ edges and $n=m/2^4$ vertices on $P$ PEs using the $G(n,m)$ generators.}
\end{figure}

\begin{figure}[t]
  \centering
  \includegraphics[width=0.49\textwidth]{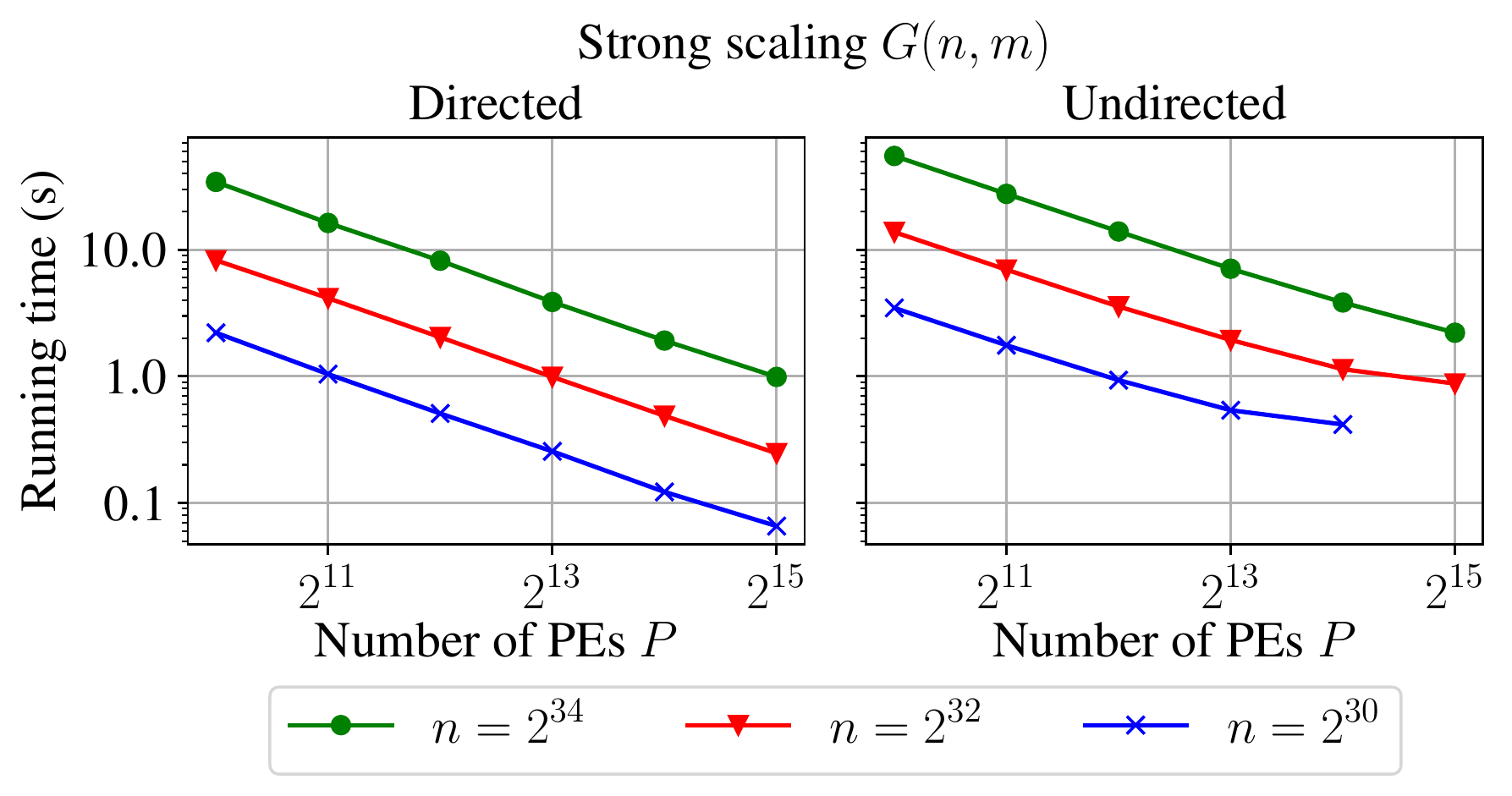}
  \vspace*{-.7cm}
  \caption{\label{fig:gnmParStrong}Running time for generating $m$ edges and $n=m/2^4$ vertices on $P$ PEs using the $G(n,m)$ generators.}
\end{figure}

Next, we discuss the scaling behavior of our \erdos~generators.
For the weak scaling experiments, each PE is assigned an equal number of $n/P$ vertices and $m/P$ edges to sample.
In particular, we set $n = m/2^{4}$ and let the number of edges per PE range from $2^{22}$ to $2^{26}$.
For the strong scaling experiments, we keep the number of edges fixed from $2^{34}$ to $2^{38}$.
Results are presented in Fig.~\ref{fig:gnmPar} and Fig.~\ref{fig:gnmParStrong} respectively.

We can see that our directed generator has an almost perfect scaling behavior.
Only for the smaller input sizes and more than $2^{12}$ PEs, the logarithmic term of our running time becomes noticeable.
The minor irregularities that we observe for the largest number of PEs are due to performance differences for nodes in the supercomputer.
Nonetheless, our results are consistent with our asymptotic running time~${\BigO((n+m)/{P} + \log P)}$.

If we look at the scaling behavior of our undirected generator, we can see that for small numbers of PEs the running time starts to increase and then remains nearly constant.
This is due to the fact that as the number of PEs/chunks increases, the number of redundantly generated edges also increases up to twice the number of sequentially sampled edges.
This effect is not noticable in the strong scaling case, since we perform these experiments with a minimum of 1024 PEs.
Furthermore, for smaller values of $m/P$ and large $P$, we also see a linear increase in running time.
We attribute this to the linear time $\BigO(P)$ needed to locate the correct chunks for each PE.

\subsection{RGG Generator}
There are various implementations of the na\"ive $\Theta(n^2)$ generator available (\eg~\cite{networkx}).
However, a more efficient and distributed algorithm is presented by Holtgrewe \etal~\cite{holtgrewe2009scalable}.

\begin{figure}[t]
  \centering
  \includegraphics[width=0.48\textwidth]{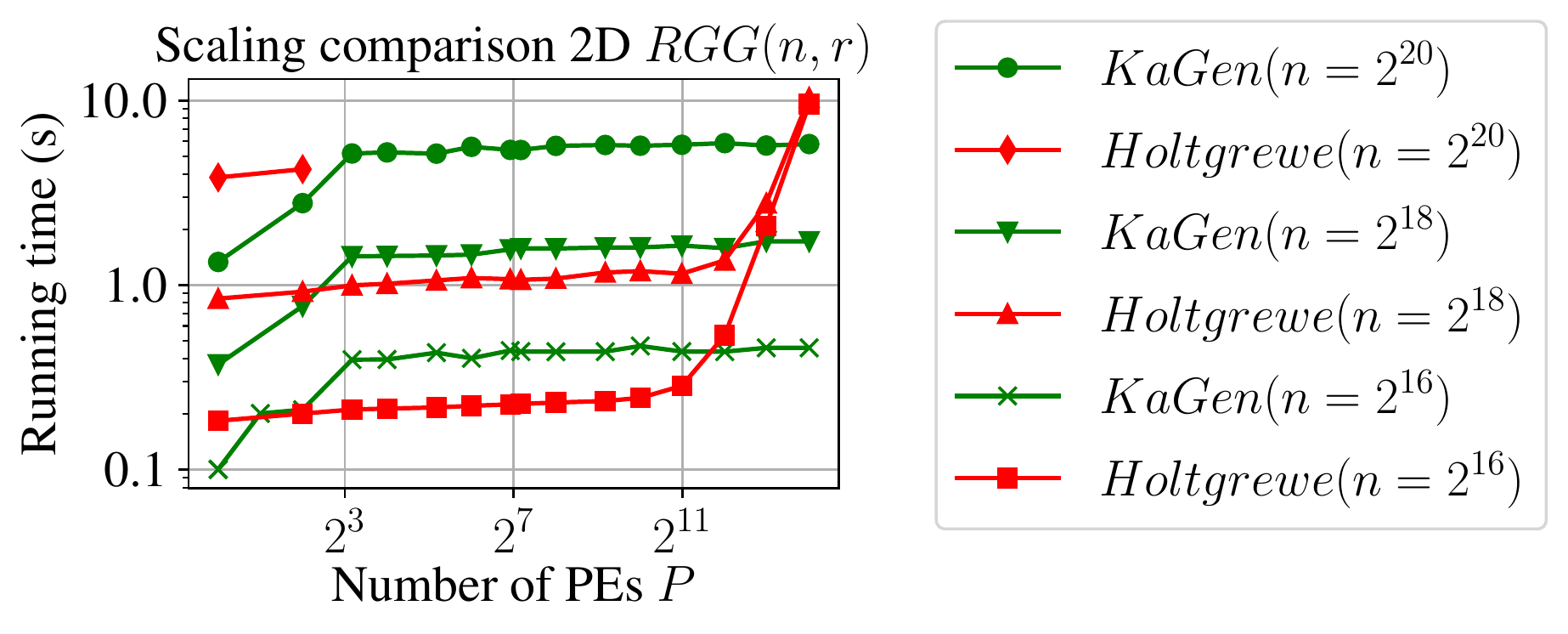}
  \vspace*{-.7cm}
    \caption{\label{fig:rgg_par} Running time for the two dimensional random geometric graph generators for growing numbers of PEs $P~=~p^2$ and a constant input size $n/P$ per PE.
    The radius is set to $r = 0.55 \sqrt{\frac{\ln n}{n}}/\sqrt{P}$.}
\end{figure}

\begin{figure}[t]
  \centering
  \includegraphics[width=0.49\textwidth]{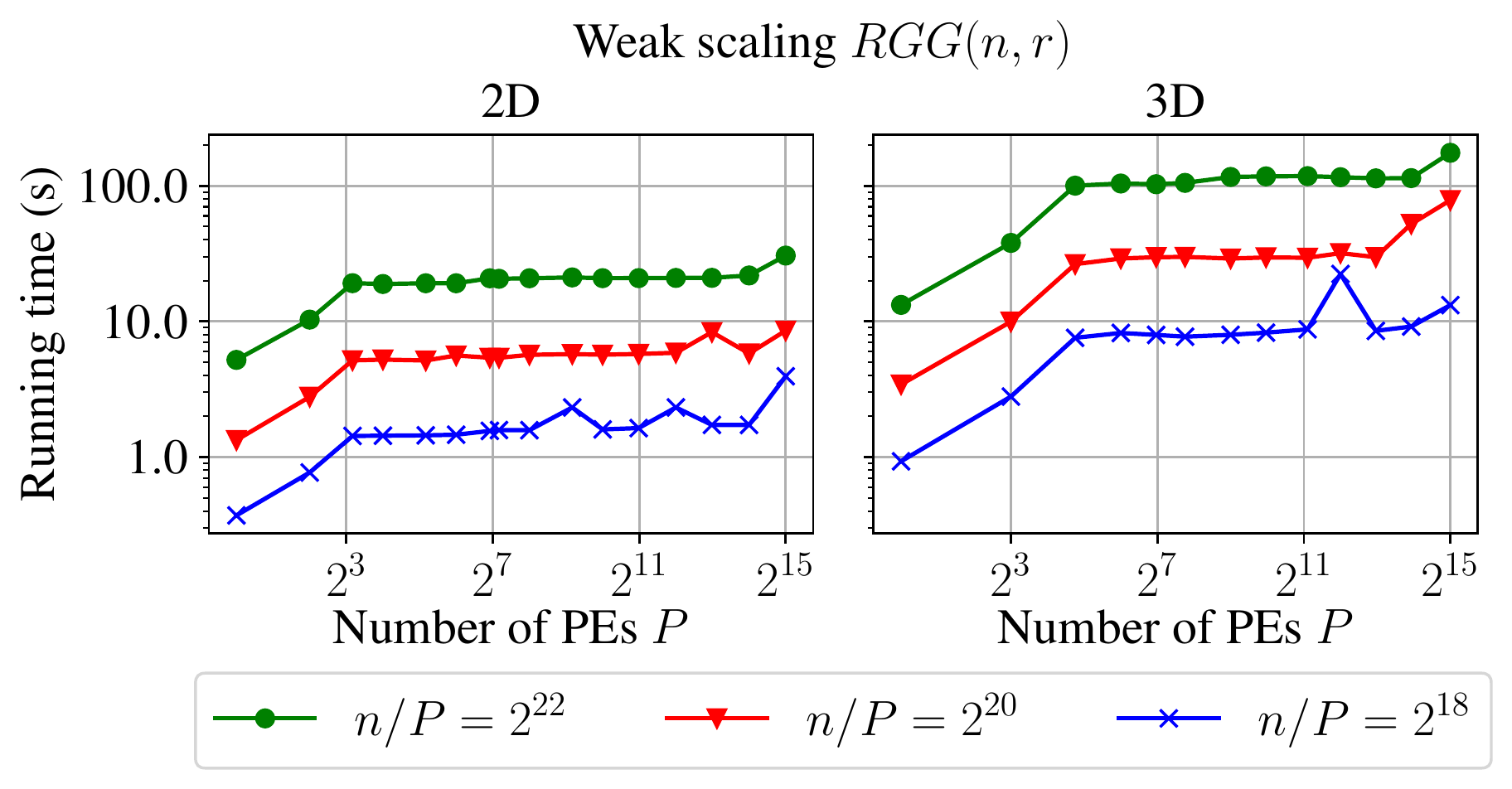}
  \vspace*{-.7cm}
    \caption{\label{fig:rggPar}Running time for generating $n$ vertices on $P$ PEs using the RGG generators. 
    The radius $r$ is set to $0.55 \sqrt[\{2,3\}]{\frac{\ln n}{n}}/\sqrt{P}$.}
\end{figure}

\begin{figure}[t]
  \centering
  \includegraphics[width=0.49\textwidth]{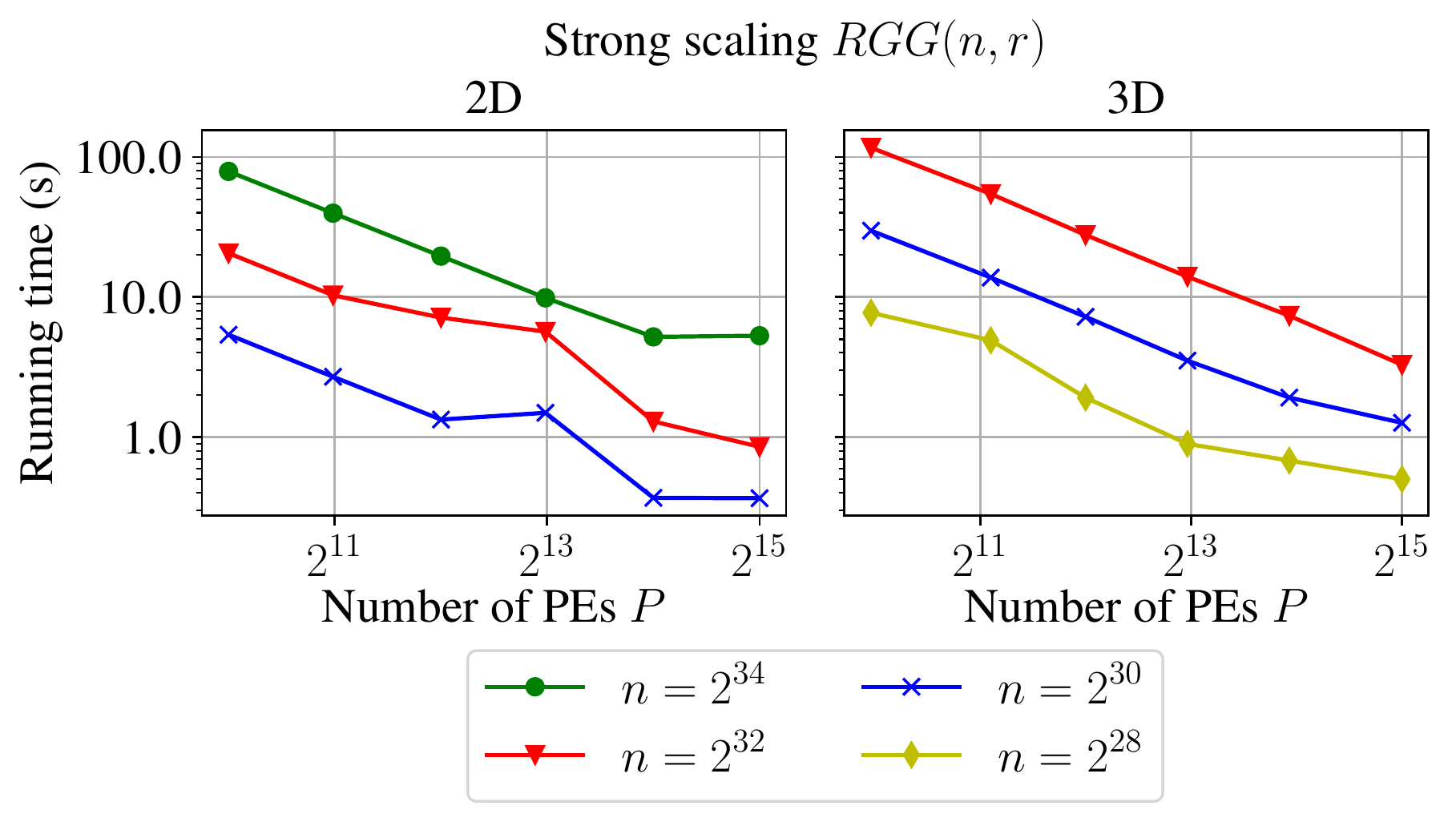}
  \vspace*{-.7cm}
    \caption{\label{fig:rggParStrong}Running time for generating $n$ vertices on $P$ PEs using the RGG generators. 
    The radius $r$ is set to $0.55 \sqrt[\{2,3\}]{\frac{\ln n}{n}}$.}
\end{figure}

Since their algorithm and our own generator are nearly identical in the sequential case, we are mainly interested in their parallel running time for a growing number of PEs.
Therefore, we measure the total running time and vary the input size per PE $n/P$ from $2^{16}$ to $2^{20}$.
It should be noted that Holtgrewe~\etal~only support two dimensional random geometric graphs and thus the three dimensional generator is excluded.
The radius is set to $r = 0.55 \sqrt{\frac{\ln n}{n}}/\sqrt{P}$.
This choice ensures that the resulting graph is almost always connected~\cite{appel2002connectivity} and is used in many previous papers.
Fig.~\ref{fig:rgg_par} shows the running time of both competitors for a growing number of $P=p^2$ PEs. 
Additionally, Fig.~\ref{fig:rggPar} shows weak scaling experiments for our two and three dimensional~generators.
Finally, we present the strong scaling behavior of our generators in Fig.~\ref{fig:rggParStrong}.
For these experiments, the number of vertices $n$ is fixed from $2^{26}$ to $2^{32}$.


Due to the recomputations used by our generator, Holtgrewe~\etal~quickly become faster by up to a factor of two as the number of PEs increases.
To be more specific, the number of neighbors that we have to generate redundantly increases from zero for one PE up to eight neighbors for more than four PEs.
This increase in running time can be bounded by computing the additional amount of vertices created through redundant computations and multiplying it by the average degree $n\pi r^2$.
For our particular choice of $r$ this yields roughly twice the running time needed for the sequential computation, which is consistent with the experimental results.
However, for sufficiently sparse graphs, the additional time for recomputations is negligible as the number of vertices in each cell becomes constant.
A very similar analysis can also be done for our three dimensional generator.
Again, minor irregularities are due to performance differences for individual nodes in the supercomputer.

Once we reach $2^{12}$ PEs, the communication required by Holtgrewe~\etal~rapidly becomes noticeable and KaGen is significantly faster.
Overall, the results are in line with the asymptotic running time presented in Section~\ref{subsec:rgg_part}.

\subsection{RDG Generator}\label{sec:eval:rdg}
Our implementation uses the CGAL library \cite{cgal} to compute the DT of the vertices of a chunk and its halo.
CGAL provides a state-of-the art implementation, 
which most of the other available DT generators use as backend as well.
We therefore omit sequential measurements.

\begin{figure}[t]
  \centering
  \includegraphics[width=0.49\textwidth]{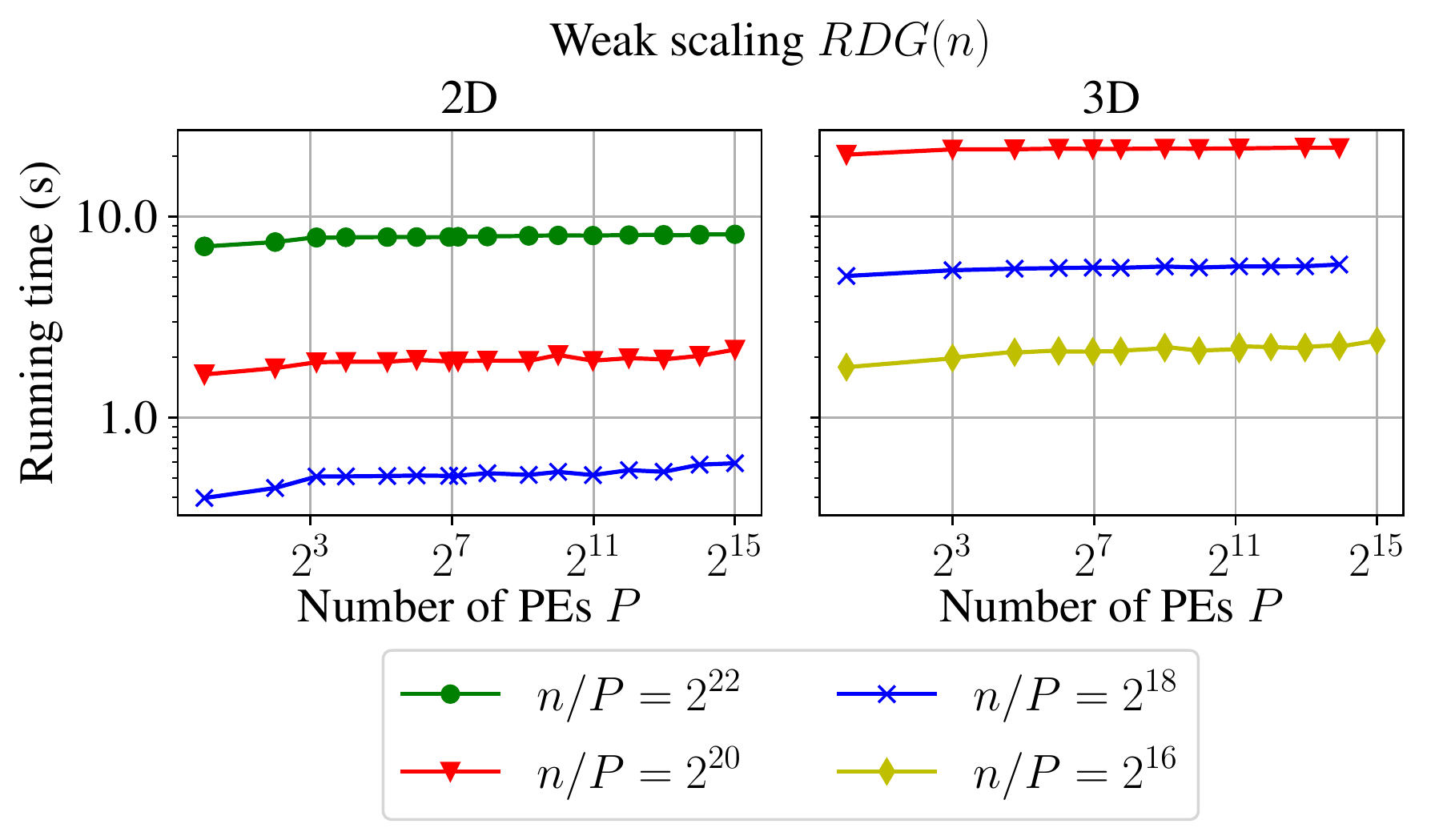}
  \vspace*{-.7cm}
    \caption{\label{fig:rdgPar}Running time for generating a graph with $n$ vertices on $P$ PEs using the RDG generators.}
\end{figure}

\begin{figure}[t]
  \centering
  \includegraphics[width=0.49\textwidth]{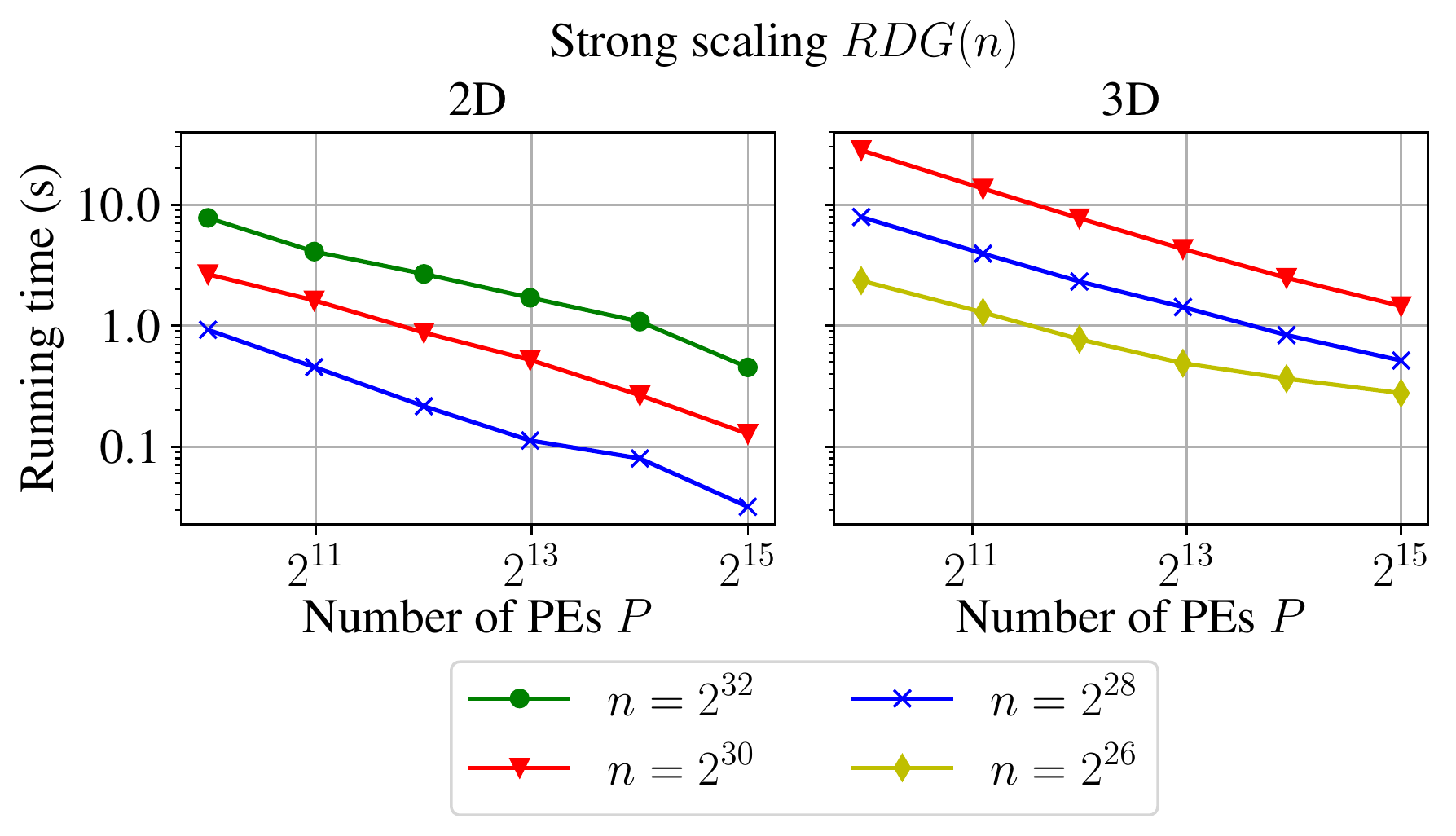}
  \vspace*{-.7cm}
    \caption{\label{fig:rdgParStrong}Running time for generating a graph with $n$ vertices on $P$ PEs using the RDG generators.}
\end{figure}

The experimental setup for the RDG is equivalent to the RGG scaling experiments.
For the weak scaling experiments, we vary the input size per PE from $2^{18}$ to $2^{22}$ for the 2D RDG and
-- due to memory constraints -- from $2^{16}$ to $2^{20}$ for the three dimensional one.
Moreover, for 3D RDG and $2^{15}$ PEs,
only the smallest input size could be computed within the memory limit per core of SuperMUC.
For the strong scaling experiments, the input size varies from $2^{26}$ to $2^{32}$.
Our experiments show an almost constant time -- depicted in Fig.~\ref{fig:rdgPar} and Fig.~\ref{fig:rdgParStrong} --
well in agreement with our conjectured asymptotic running time of 
$\BigO({n}/{P} + \log P)$.
Similarly to the RGG, the initial increase in runtime
can be attributed to the redundant vertex generation of neighboring cells.
As the halo rarely grows beyond the directly adjacent cells, no significant further increase in runtime can be observed for more than
$2^8$ PEs.

\subsection{RHG Generator}
\begin{figure}
    \centering
    \includegraphics[width=0.49\textwidth]{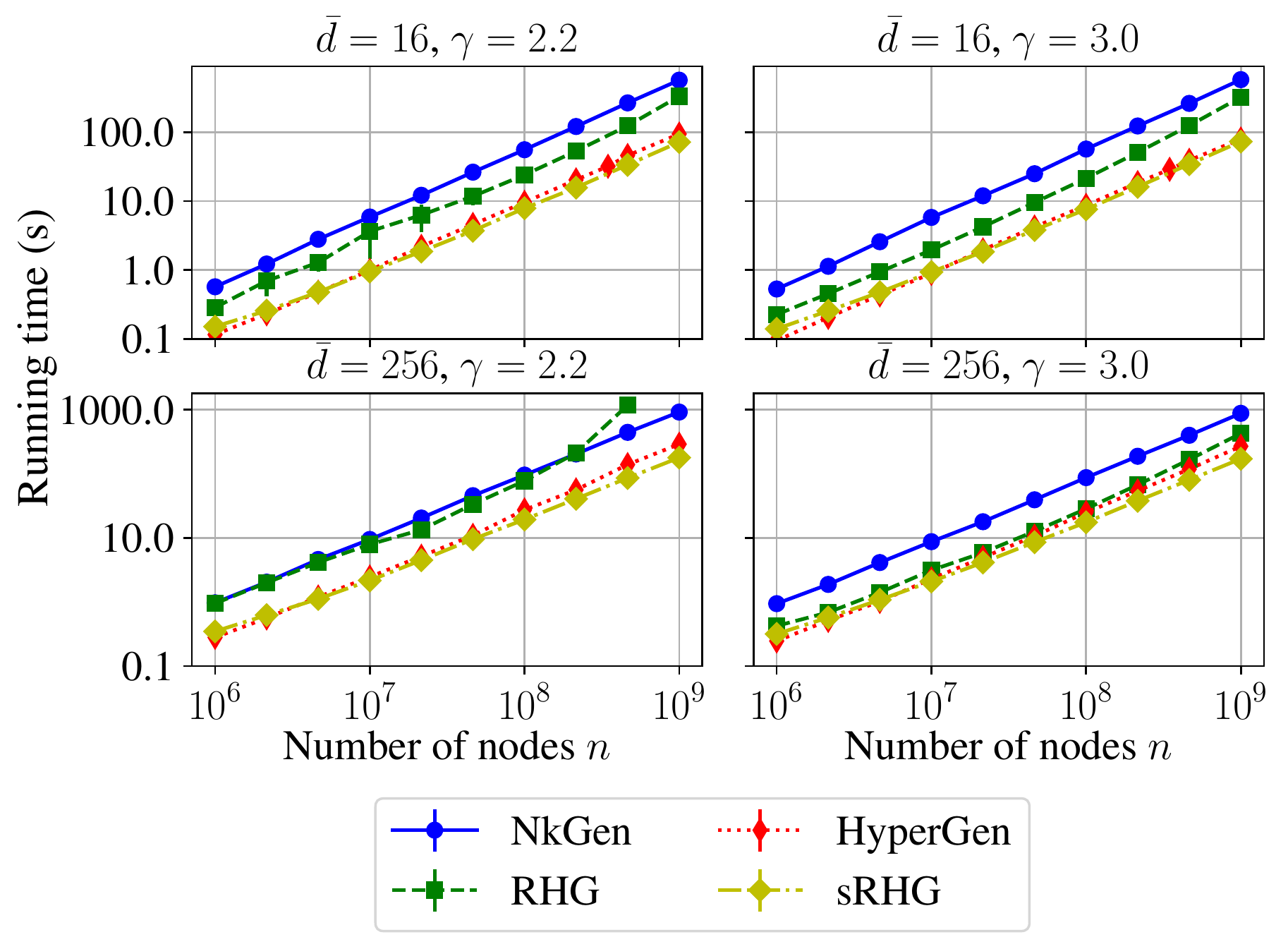}
    \caption{
        Running time as function of number $n$ of nodes for power-law exponents $\gamma \in \{2.2, 3\}$ (i.e., $\alpha \in \{0.6, 1\}$) and average degrees $\bar d \in \{16, 256\}$.
        All generators use 39 threads or processes, on a dual-socket Intel Xeon Broadwell E5-2640~v4 machine with and 128~GB of RAM.
    }
    \label{fig:rhg_compare}
\end{figure}

We compare RHG and sRHG to the state-of-the-art generators NkGen\footnote{
    Optimized version of the generator found in the NetworKit library \url{http://network-analysis.info}~\cite{penschuck2017generating}.
} by von Looz~\etal~\cite{von2016generating}, and HyperGen by Penschuck~\cite{penschuck2017generating} (see section~\ref{subsec:related_rhg}).
Since both reference implementations support shared-memory parallelism only, we restrict the experiments to a single machine with 40~hardware threads.

We measure the runtime of all generators as the number $n$ of nodes varies between $10^6 \le n \le 10^9$ for different average degrees $\bar d$ and power-law exponents~$\gamma$.
These values mimic settings found in various real-world networks~\cite{von2015generating}.

In general, NkGen exhibits the highest runtime per edge generated.
We attribute this to the fact that the implementation uses only partial pre-computation and heavily relies on unstructured accesses to main memory.
NkGen is typically followed by RHG which demonstrates limited scaling for small values of $\gamma$;
in this setting, the increase in runtime for large graphs is primarily caused by exhaustion of the system's main memory.
We consider this unrepresentative for the distributed case, in which the collective memory available is typically much larger.

The related sRHG and HyperGen are consistently the fastest implementations with sRHG producing up~to $7.5 \cdot 10^8$ edges per second for $\bar d=256$ and $\gamma = 3$ on a single machine.
Similar to RHG both use pre-computation as a means to speed up distance computation.
Additional performance gains are due their emphasis on cache and memory~efficiency and data parallelism.

\begin{figure}[t]
  \centering
  \includegraphics[width=0.49\textwidth]{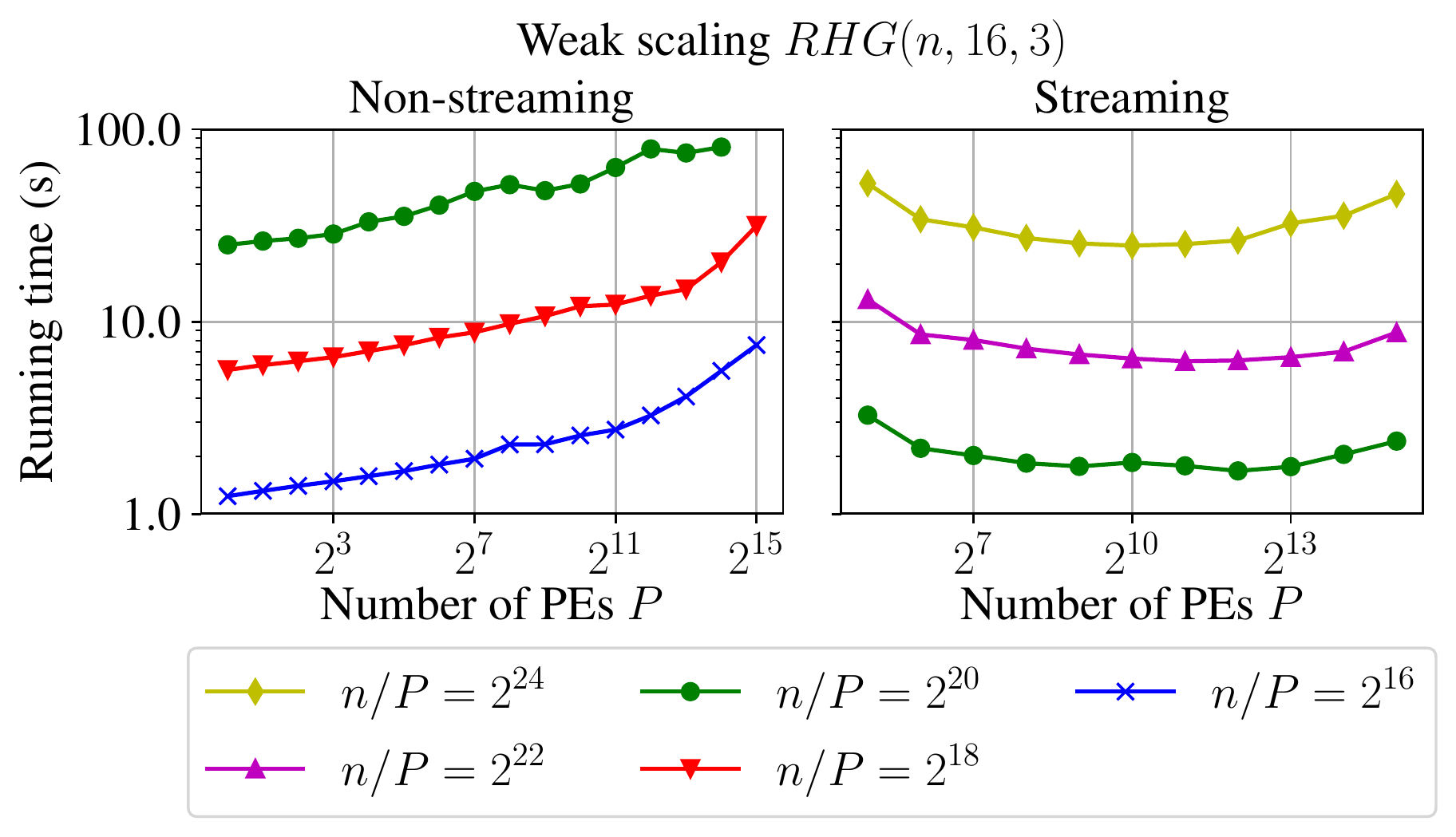}
  \vspace*{-.7cm}
  \caption{\label{fig:rhgPar3}Running time for generating a graph with $n$ vertices, average degree $\bar{d}=16$ and $\gamma=3.0$ on $P$ PEs using both RHG generators.}
\end{figure}

\begin{figure}[t]
  \centering
  \includegraphics[width=0.49\textwidth]{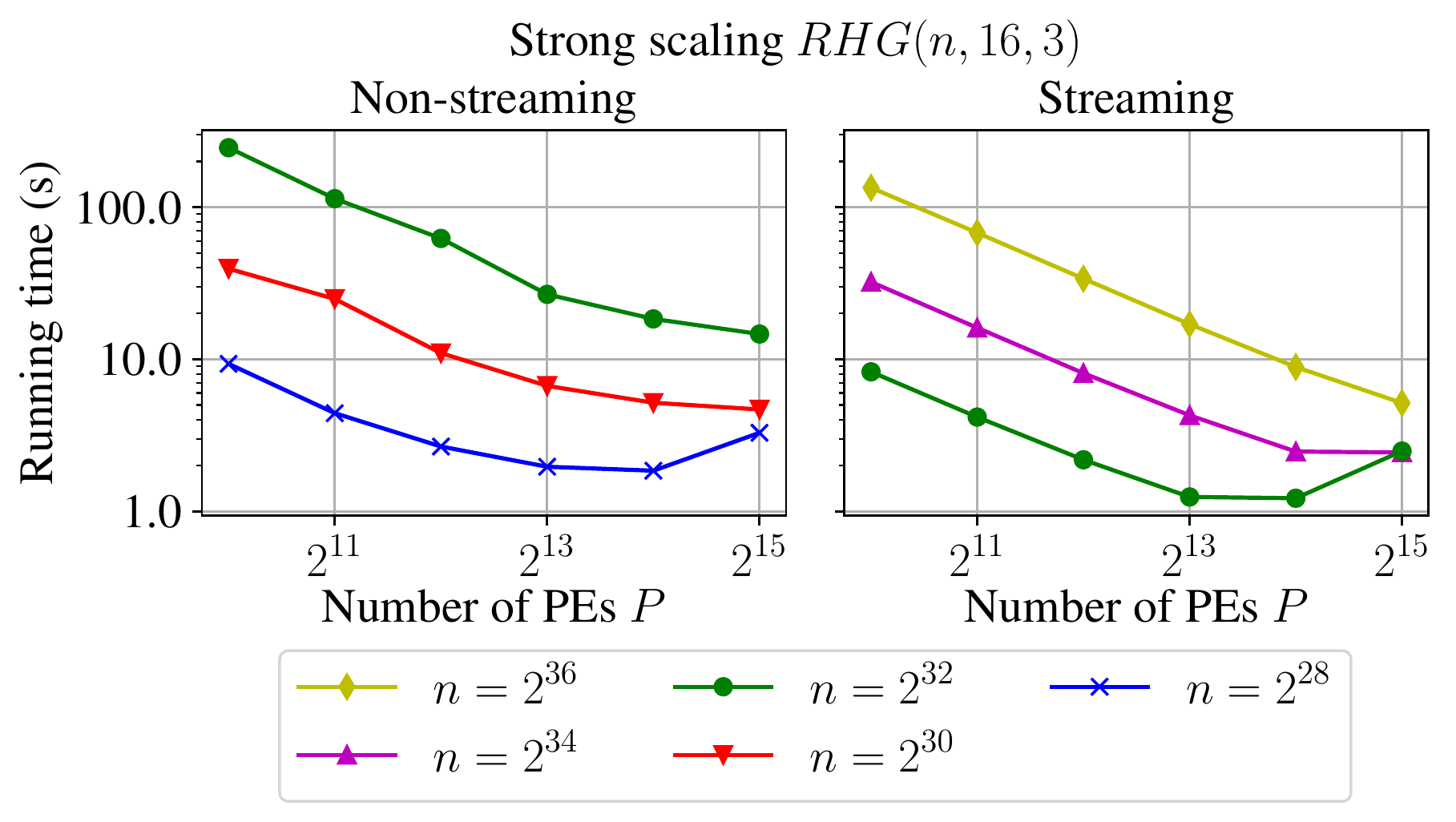}
  \vspace*{-.7cm}
  \caption{\label{fig:rhgPar3Strong}Running time for generating a graph with $n$ vertices, average degree $\bar{d}=16$ and $\gamma=3$ on $P$ PEs using the non-streaming RHG generator.}
\end{figure}

Finally, we present the results of our scaling experiments for the RHG generators.
For the weak scaling experiments each PE is again assigned an equal number of $n$ vertices.
Note, that we use a designated computing node with 16 cores for calculating the inner core for our second generator.
Thus, the corresponding scaling experiments start at 32 cores.
The number of vertices per PE ranges from $2^{16}$ to $2^{24}$.
Again, for the strong scaling experiments the number of vertices is fixed and ranges from $2^{28}$ to $2^{32}$.
The power-law exponent $\gamma$ varies from $2.2$ to $3.0$ to cover different extremes of the degree distribution.
Fig.~\ref{fig:rhgPar3} ($\gamma=3$) shows the weak scaling results of the RHG generators with an average degree of $\bar{d} = 16$.
Likewise, Fig.~\ref{fig:rhgPar3Strong} ($\gamma=3$) shows the strong scaling results of the RHG generators with an average degree of $\bar{d} = 16$.

Looking at the scaling behavior of our first generator, we see that there is a considerable increase in running time for a growing number of PEs.
We can attribute this behavior to the redundant computations that are introduced through parallelization.
Additionally, high degree vertices are hard to distribute efficiently if we want a partitioned output graph. This severely impedes the scaling behavior.
Since the maximum degree is $\BigO(n^{\frac 1 {2 \alpha}})$ with high probability~\cite{gugelmann2012random}, these vertices dominate the running time of our algorithm.
Overall, these effects are less noticable in the case of our strong scaling experiments, because we start with a minimum of 1024 PEs.

If we examine the scaling behavior of our second algorithm, we can see that these effects are less noticeable, especially for larger values of $\gamma$.
For smaller values of $\gamma$, the running time of our algorithm is again dominated by the time needed to generate the inner core (global annuli) on its dedicated computing node.
Overall, our second generator is roughly $16$ times faster than our first generator.
However, keep in mind that the resulting graph is not fully partitioned, \ie~not all incident edges are generated on the corresponding PEs.
Thus, we can achieve a similar speedup for our first generator, by only performing outward queries and omitting the inward ones.
Nonetheless, the lower memory requirements of our second generator enable us to generate up to $16$ times larger instances.


\subsubsection{Comparison with R-MAT}
\begin{figure}[t]
  \centering
  \includegraphics[width=0.43\textwidth]{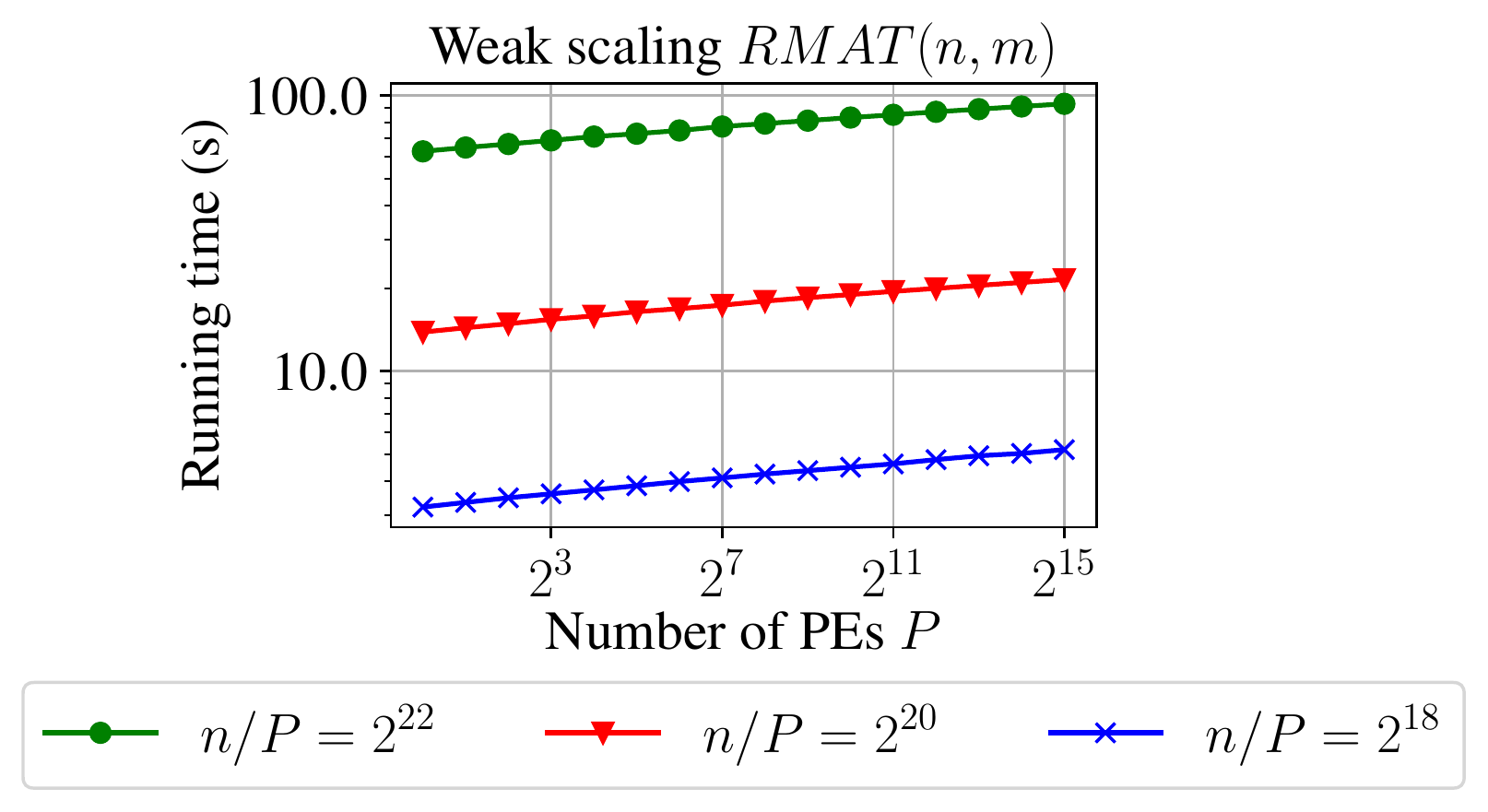}
  \caption{\label{fig:rmatPar}Running time for generating a graph with $n$ vertices and $m=2^4n$ edges on $P$ PEs using the R-MAT generator.}
\end{figure}

In order to further evaluate the performance of our generators, in particular of our hyperbolic generators, we now compare them to the R-MAT generator. 
For this purpose, we use the reference implementation available at the Graph~500 website (\url{graph500.org}).
The R-MAT generator is commonly used in benchmarks for large scale graph computations due to its scalability and flexibility.
Fig.~\ref{fig:rmatPar} shows the weak scaling behavior of R-MAT an equal number of $m/P$ edges per PE.
In particular, we set $n = m/2^{4}$ and let the number of edges per PE range from $2^{22}$ to $2^{26}$.

First, we see that R-MAT has a slight increase in running time for growing numbers of PEs (and thus a growing number of vertices).
This increase in running time is due to the fact that R-MAT needs to generate $\BigO(\log n)$ random variates for each edges. 
Second, if we compare the overall performance of R-MAT with our own generators we can see that it is roughly ten times slower than the streaming version of our hyperbolic graph generator.
Furthermore, it is up to 15 times slower than our undirected \erdos~generator.
This difference in performance can be attributed to the difference in the number of random variates that each generator has to generate.
Due to the costs of generating these variates, minimizing the number of variates yields large performance benefits.

\begin{figure}[t]
  \centering
  \includegraphics[width=0.37\textwidth]{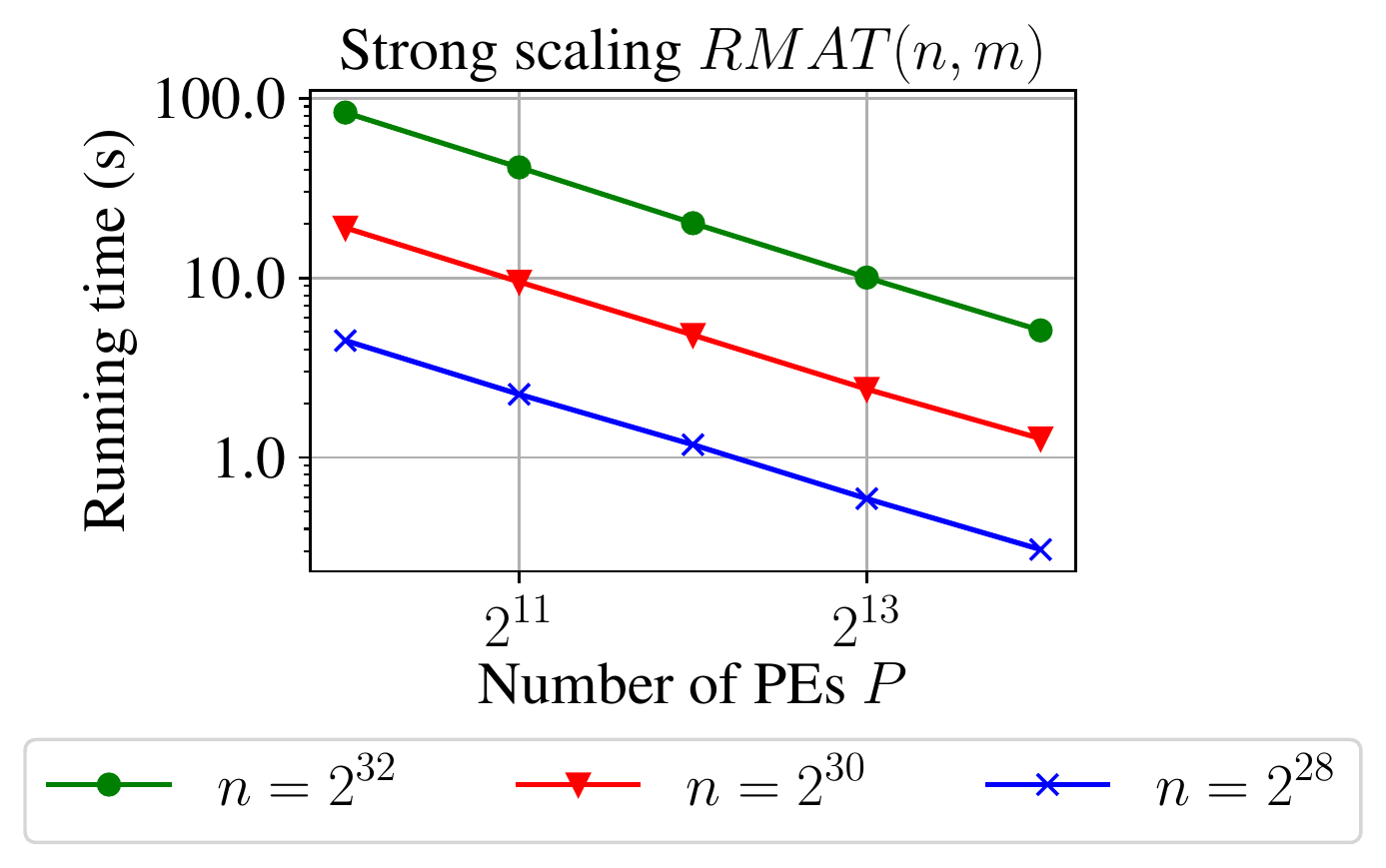}
  \caption{\label{fig:rmatParStrong}Running time for generating a graph with $n$ vertices and $m=2^4n$ edges on $P$ PEs using the R-MAT generator.}
\end{figure}

\section{Conclusion}
We presented scalable graph generators for a set of commonly used network models.
Our work includes the classic \erdos~model, in both the $G(n,m)$ and $G(n,p)$ variants, random geometric graphs, random Delaunay graphs and random hyperbolic graphs.

Our algorithms make use of a combination of divide-and-conquer schemes and grid data structures to narrow down their local sampling space.
We redundantly compute parts of the network that are within the neighborhood of local vertices. 
These computations are made possible through the use of pseudorandomization via hash functions.
The resulting algorithms are embarrassingly parallel and \emph{communication-free}.

Our extensive experimental evaluation indicates that our generators are competitive to state-of-the-art algorithms while also providing near-optimal scaling behavior. 
In turn, we are able to generate instances with up to $2^{43}$ vertices and $2^{47}$ edges in less than 22 minutes on 32\,768 cores.\footnote{Using the directed $G(n,m)$ generator.}
Therefore, our generators enable new network models to be used for research on a massive scale.
In order to help researchers to use our generators, we provide all algorithms in a widely usable open-source library.
Finally, to show the broad applicability of the concepts used in our generators, we provide adaptations for their use in a GPGPU setting.

\subsubsection*{Future Work}
\label{sec:future}
As mentioned in Section~\ref{sec:exp_setup}, we would like to extend our remaining generators to use a streaming approach similar to sRHG (see Section~\ref{subsec:srhg}).
This would drastically reduce the memory needed for the auxiliary data structures, especially for the spatial network generators.
Allowing the efficient generation of random hyperbolic graphs on GPGPUs also remains for future work.

Furthermore, we would like to extend our communication-free paradigm to various other network models such as the stochastic block-model and the probabilistic random hyperbolic graph model~\cite{papadopoulos2008greedy}.
More specifically, we would like to extend the KaGen library by a faster generator for R-MAT graphs than currently available.

Finally, our generators allow us to perform an extensive study on new graph models for high-performance computing benchmarks.
In turn, these benchmarks could target a wider variety or real-world models and scenarios.
A more detailed theoretical analysis using tighter bounds, especially for the parallel running times of our generators would be beneficial for this purpose.


\section*{Acknowledgment}
The authors gratefully acknowledge the Gauss Centre for Supercomputing e.V. (\url{www.gauss-centre.eu}) for funding this project by providing computing time on the GCS Supercomputer SuperMUC at Leibniz Supercomputing Centre (\url{www.lrz.de}).
The authors gratefully acknowledge the Gauss Centre for Supercomputing (GCS) for providing computing time through the John von Neumann Institute for Computing (NIC) on the GCS share of the supercomputer JUQUEEN~\cite{stephan2015juqueen} at J{\"u}lich Supercomputing Centre (JSC). GCS is the alliance of the three national supercomputing centres HLRS (Universit\"at Stuttgart), JSC (Forschungszentrum J{\"u}lich), and LRZ (Bayerische Akademie der Wissenschaften), funded by the German Federal Ministry of Education and Research (BMBF) and the German State Ministries for Research of Baden-W{\"u}rttemberg (MWK), Bayern (StMWFK) and Nordrhein-Westfalen (MIWF).
We thank the Center for Scientific Computing, University of Frankfurt for making their HPC facilities available.
This work was partially supported by Deutsche Forschungsgemeinschaft (DFG) under grants ME~2088/3-2, and ME~2088/4-2.

\appendix
\section{Hyperbolic Geometry Related Definitions}
\noindent Radial density:
\begin{align}
  \rho(r) &\defrel \alpha \frac{\sinh(\alpha r)}{\cosh(\alpha R) - 1}
  \intertext{Radial cdf:}
	\mu(B_r(0)) &\defrel \int_{0}^{r} \rho(x) \intd x = \frac{\cosh(\alpha x) - 1}{\cosh(\alpha R)} 
  \intertext{Angular deviation:}\label{eq:deltatheta}
	\Delta\theta(r, b) &\defrel 
	\begin{cases}
		\pi & \text{if } r{+}b < R \\
		\arccos\big[\frac{\cosh(r)\cosh(b) - \cosh(R)}{\sinh(r)\sinh(b)}\big] & \text{otherwise}
	\end{cases}
\end{align}

\vspace{1em}

\section{Hyperbolic Geometry Related Approximations}\label{subsec:def-approx}
Gugelmann et al. derived the following approximations\footnote{We drop the $(1 + \BigO(\cdot))$ error terms in our calculations without further notice if they are either irrelevant or dominated by other simplifications made}~\cite{gugelmann2012random}.\\

\noindent Angular deviation:
\begin{align}\label{eq:deltathetaapprox}
	\Delta\theta(r, b) &= 
	\begin{cases}
	\pi & \text{if } r+b < R \\
	2\e^\frac{R-r-b}{2} (1 + \Theta(e^{R-r-b})) & \text{if } r+b \ge R
	\end{cases}\\	
\intertext{Radial cdf:}\label{eq:cdf}
	\mu(B_r(0)) 
		&= \int^{r}_{0} \rho(x) \intd x 
    = \frac{\cosh(\alpha r)}{\cosh(\alpha R) - 1} \\
    &= \e^{\alpha(r-R)} (1 + o(1)) 
\end{align}

The probability mass $\mu_Q$ of the intersection of the actual query circle $B_R(r)$ with the annulus $B_b(0) \setminus B_a(0)$ as defined in Lemma~\ref{lem:candidates} is given by:
\begin{align}
\mu_Q &\defrel \mu\left[ (B_b(0) {\setminus} B_a(0)) \cap B_R(r) \right] \nonumber\\
&\stackrel{\phantom{}}{=} \frac 2 \pi \e^{-\frac r 2 - (\alpha-\frac 12)R} 
\Bigg[
\frac{\alpha}{\alpha{-}\frac 1 2}\left(
\e^{(\alpha-\frac 1 2)b} - \e^{(\alpha-\frac 1 2)a}
\right)\nonumber \\ 
& \hspace{3em} + \errorterm{\BigO\left(\e^{-(\alpha{-}\frac 1 2)a}\right)}
\Bigg] \label{eq:mu_actual_query}
\end{align}

RHG and sRHG overestimate the actual query range at the border and covers the mass $\mu_H$:
\begin{align}
\mu_H &\defrel \frac 1 \pi \Delta\theta(r, a) \int\limits_a^b \rho(y) dy \nonumber \\
&=\frac 2 \pi \e^{-\frac r 2 - (\alpha-\frac 12)R} \left[ 
\e^{\alpha b - a/2} - \e^{(\alpha - \frac 12)a}
\right] \nonumber\\
& \hspace{3em} \cdot \errorterm{\left(1 \pm \BigO\left(\e^{(1-\alpha)(R-a)-r}\right)\right)}
\label{eq:mu_overestimated}
\end{align}

\bibliographystyle{elsarticle-num} 
\bibliography{extracted}

\end{document}